\documentclass[aps,twocolumn,,eqsecnum,prb]{revtex4}
\usepackage[dvips]{graphicx}
\usepackage{amsmath,amsbsy,amssymb,graphics}
\usepackage[mathscr]{euscript}

\def\beginABC{\begin{subequations}}
\def\endABC{\end{subequations}}
\let\mathit=\mathscr
\let\mathbf=\boldsymbol

\begin{document}

\title{Ground-State Structure in $\nu =2$ Bilayer Quantum Hall Systems}
\author{Z.F. Ezawa$^{1}$, M. Eliashvili$^{2}$ and G. Tsitsishvili$^{1,2}$}
\affiliation{{}$^1$Department of Physics, Tohoku University, Sendai, 980-8578 Japan }
\affiliation{{}$^2$Department of Theoretical Physics, A. Razmadze Mathematical Institute,
Tbilisi, 380093 Georgia}
\date{\today}

\begin{abstract}
We investigate the ground-state structure of the bilayer quantum Hall system
at the filling factor $\nu =2$. Making an exact analysis of the ground state
in the SU(4)-invariant limit, we include all other interactions as small
perturbation. We carry out analytic calculations and construct phase
diagrams for nonzero values of the Zeeman, tunneling and bias interactions.
In particular we examine carefully how the phase transition occurs by
applying the bias voltage and inducing a density imbalance between the two
layers. We compare our theoretical result with the experimental data due to
Sawada et al. based on the phase diagram in the $\sigma _{0}$-$\rho _{0}$
plane, where $\rho _{0}$ and $\sigma _{0}$ are the total electron density
and the density difference between the two layers, respectively.
\end{abstract}

\maketitle

\section{Introduction}

A rich physics has emerged from the layer degree of freedom in the bilayer
quantum Hall (QH) system\cite{BookEzawa,BookDasSarma}. An electron carries
the SU(2) pseudospin index assigned to the front and back layers. The spin
and the pseudospin are equal partners. There arises a unique phase at the
filling factor $\nu =1$. It is the spin-ferromagnet and
pseudospin-ferromagnet phase, where various intralayer and interlayer
coherent phenomena have been observed\cite{BookEzawa,BookDasSarma}. On the
other hand, according to the one-body picture we expect to have two phases
at $\nu =2$ depending on the relative strength between the Zeeman gap $%
\Delta _{\text{Z}}$ and the tunneling gap $\Delta _{\text{SAS}}$. One is the
spin-ferromagnet and pseudospin-singlet phase (abridged as the spin phase%
\cite{NoteA}) for $\Delta _{\text{Z}}>\Delta _{\text{SAS}}$; the other is
the spin-singlet and pseudospin-ferromagnet phase (abridged as the ppin phase%
\cite{NoteA}) for $\Delta _{\text{Z}}<\Delta _{\text{SAS}}$. Between these
two phases, driven by interlayer correlations, a novel canted
antiferromagnetic phase has been predicted to emerge\cite{Zheng97L,Sarma97L}.

The first experimental indication of a phase transition at $\nu =2$ was
revealed by inelastic light scattering spectroscopy\cite{Pellegrini97L}. An
unambiguous evidence was subsequently obtained through magneto-transport
measurements\cite{Sawada98L}, where the importance of making density
imbalance between the two layers is emphasized to study phase transitions.
There is an attempt\cite{Brey99L} to interpret the data\cite{Sawada98L}
based on numerical analyses. However, the physical picture is not yet clear
enough due to the lack of analytic understanding. For instance, it is yet to
be explored how the density imbalance is induced as a function of the bias
voltage $V_{\text{bias}}$. The density imbalance is made as soon as $V_{%
\text{bias}}\neq 0$ at $\nu =1$, but this is not so simple at $\nu =2$.
Furthermore, the exact diagonalization\cite{Schliemann00L} of a few-electron
system has shown that the boundary between the spin and canted phases is
practically unmodified but the boundary between the canted and ppin phases
is considerably modified from the mean-field result. There is so far no
theoretical explanation of these behaviors. The aim of this paper is to
understand the ground-state structure of the $\nu =2$ bilayer QH system more
in details based on perturbation theory.

We investigate physics taking place in the lowest Landau level (LLL). Since
each Landau site can accommodate four electrons with the spin and pseudospin
degrees of freedom, the underlying group structure is enlarged to SU(4). We
start with the SU(4)-invariant limit of the bilayer system, where the ground
state is determined by solving the eigenvalue equation. We then include
SU(4)-noninvariant terms perturbatively. The ground-state energy is
calculated analytically in the first order of perturbation. Then we examine
carefully how the phase transition occurs as the density imbalance is made
between the two layers. We are particularly interested in the phase diagram
in the $\sigma _{0}$-$\rho _{0}$ plane, where $\rho _{0}$ and $\sigma _{0}$
denotes the total electron density and the imbalance parameter (the
normalized density difference between the two layers), respectively. This is
because phase transition points have experimentally\cite{Sawada98L} been
observed most clearly in such a plane. We also present the phase diagram in
the the $\rho _{0}$-$\Delta _{\text{SAS}}$ plane, which will be useful to
analyze experimental data obtained in samples with different $\Delta _{\text{%
SAS}}$ but all other parameters unchanged.

In Section II we review the Landau-site Hamiltonian governing bilayer QH
systems\cite{EzawaX04B}. In Section III we solve the eigenvalue equation for
the ground state in the SU(4)-invariant limit. In so doing we identify the
physical variables within 15 generators of the group SU(4). We show that
there are 9 independent physical variables, among which 8 variables describe
real Goldstone modes on the SU(4)-invariant ground state. We then calculate
the ground-state energy of the full system in the first order of
perturbation. In Section IV we verify the existence of the spin phase, the
canted phase and the ppin phase in the presence of the bias voltage. The
ground state is obtained in an analytic form in each phase. In particular
the spin component $S_{z}$ and the imbalance parameter $\sigma _{0}$ are
calculated as functions of the bias voltage for typical sample parameters.
In Section V we present phase diagrams in the $\Delta _{\text{SAS}}$-$\Delta
_{\text{Z}}$ plane, in the $\sigma _{0}$-$\rho _{0}$ plane and in the $\rho
_{0}$-$\Delta _{\text{SAS}}$ plane. In Section VI we show that the spin
phase is not modified but the ppin phase is considerably modified by higher
order perturbations. Thus, the spin-canted phase boundary is an exact
result, as is consistent with the result\cite{Schliemann00L} obtained from
the exact diagonalization of a few-electron system. In Section VII we
interpret the experimental data\cite{Sawada98L} based on the phase diagram
in the $\sigma _{0}$-$\rho _{0}$ plane. Section VIII is devoted to
discussions.

\section{Landau-site Hamiltonian}

\label{SecLandaHamil}

Electrons in a plane perform cyclotron motion under strong magnetic field $%
B_{\perp }$ and create Landau levels. In QH systems the electron position is
specified solely by the guiding center $\mathbf{X}=(X,Y)$ subject to the
noncommutative relation, $[X,Y]=-i\ell _{B}^{2}$, with the magnetic length $%
\ell _{B}=\sqrt{\hbar /eB_{\perp }}$. It follows from this relation that
each electron occupies an area $2\pi \ell _{B}^{2}$ labelled by the
Landau-site index. The total number of Landaus sites is $N_{\Phi }=S/2\pi
\ell _{B}^{2}=N/\nu $, where $N$ and $S$ are the total electron number and
the area of the system, respectively. Electrons behave as if they were on
lattice sites, among which Coulomb interactions operate.

In the bilayer system an electron carries $4$ different polarizations
associated with the ordinary spin $(\uparrow ,\downarrow )$ and two layers $%
( $f,b$)$. These four polarizations can be incorporated in the isospin index 
$\mu =$f$\uparrow $, f$\downarrow $, b$\uparrow $, b$\downarrow $. In this
notation the second quantized electron field in the lowest Landau level
(LLL) appear as 
\begin{equation}
\psi _{\mu }(\mathbf{x})=\sum_{m=0}^{\infty }c_{\mu }(m)\langle \mathbf{x}%
|m\rangle ,  \label{FieldExpan}
\end{equation}%
where $\langle \mathbf{x}|m\rangle $ is the one-body wave function for the
Landau site $|m\rangle $ in the LLL with $m$ labeling the angular momentum.
The operators $c_{\mu }(m)$ and $c_{\nu }^{\dagger }(n)$ satisfy the
standard anticommutation relations.

The electron field $\psi _{\mu }$ has four components, and the bilayer
system possesses the underlying algebra SU(4). It has the subalgebra SU$_{%
\text{spin}}$(2)$\otimes $SU$_{\text{ppin}}$(2). We denote the 3 generators
of the spin SU(2) algebra by $\tau _{a}^{\text{spin}}$, and those of the
pseudospin SU(2) generators by $\tau _{a}^{\text{ppin}}$. There are
remaining $9$ generators in SU(4), which are given by $\tau _{a}^{\text{spin}%
}\tau _{b}^{\text{ppin}}$. Their properties are summarized as 
\begin{align}
\tau _{a}^{\text{spin}}\tau _{b}^{\text{spin}}=& \delta _{ab}\mathbb{I}%
+i\varepsilon _{abc}\tau _{c}^{\text{spin}},  \notag \\
\tau _{a}^{\text{ppin}}\tau _{b}^{\text{ppin}}=& \delta _{ab}\mathbb{I}%
+i\varepsilon _{abc}\tau _{c}^{\text{ppin}},  \notag \\
\tau _{a}^{\text{spin}}\tau _{b}^{\text{ppin}}=& \tau _{b}^{\text{ppin}}\tau
_{a}^{\text{spin}},
\end{align}%
where $\mathbb{I}$ stands for the $4\times 4$ identity matrix. Their
explicit form is given in Appendix \ref{AppenPV}.

All the physical operators required for the description of the system are
constructed as bilinear combinations of $\psi (\mathbf{x})$ and $\psi ^{\dag
}(\mathbf{x})$. They are 16 density operators%
\begin{align}
\rho (\mathbf{x})=& \psi ^{\dag }(\mathbf{x})\psi (\mathbf{x}),  \notag \\
S_{a}(\mathbf{x})=& \frac{1}{2}\psi ^{\dag }(\mathbf{x})\tau _{a}^{\text{spin%
}}\psi (\mathbf{x}),  \notag \\
P_{a}(\mathbf{x})=& \frac{1}{2}\psi ^{\dag }(\mathbf{x})\tau _{a}^{\text{ppin%
}}\psi (\mathbf{x}),  \notag \\
R_{ab}(\mathbf{x})=& \frac{1}{2}\psi ^{\dag }(\mathbf{x})\tau _{a}^{\text{%
spin}}\tau _{b}^{\text{ppin}}\psi (\mathbf{x}),
\end{align}%
where $S_{a}$ describes the total spin, $2P_{z}$ measures the
electron-density difference between the two layers. The operator $R_{ab}$
transforms as a spin under SU$_{\text{spin}}$(2) and as a pseudospin under SU%
$_{\text{ppin}}$(2).

The total Hamiltonian consists of the Coulomb, Zeeman, tunneling and bias
terms. The Coulomb interaction is decomposed into the SU(4)-invariant and
SU(4)-noninvariant terms, 
\begin{align}
H_{C}^{+}=& \frac{1}{2}\int \!d^{2}xd^{2}y\,\rho (\mathbf{x})V^{+}(\mathbf{x}%
-\mathbf{y})\rho (\mathbf{y}),  \notag \\
H_{C}^{-}=& 2\int \!d^{2}xd^{2}y\,P_{z}(\mathbf{x})V^{-}(\mathbf{x}-\mathbf{y%
})P_{z}(\mathbf{y}),
\end{align}%
where 
\begin{equation}
V^{\pm }(\mathbf{x})=\frac{e^{2}}{8\pi \varepsilon }\left( \frac{1}{|\mathbf{%
x}|}\pm \frac{1}{\sqrt{|\mathbf{x}|^{2}+d^{2}}}\right)  \label{CouloPM}
\end{equation}%
with $d$ being the layer separation. The tunneling and bias terms are
summarized into the pseudo-Zeeman term. Combining the Zeeman and
pseudo-Zeeman terms we have 
\begin{equation}
H_{\text{ZpZ}}=-\int \!d^{2}x\,\left( \Delta _{\text{Z}}S_{z}+\Delta _{\text{%
SAS}}P_{x}+\Delta _{\text{bias}}P_{z}\right) ,
\end{equation}%
with the Zeeman gap $\Delta _{\text{Z}}$, the tunneling gap $\Delta _{\text{%
SAS}}$\ and the bias term $\Delta _{\text{bias}}=eV_{\text{bias}}$.

Substituting the field expansion (\ref{FieldExpan}) into $H_{C}^{\pm }$ we
obtain Landau-site Hamiltonians\cite{EzawaX04B},%
\begin{align}
H_{C}^{+}=& \sum_{mnij}V_{mnij}^{+}\rho (m,n)\rho (i,j),  \notag \\
H_{C}^{-}=& 4\sum_{mnij}V_{mnij}^{-}P_{z}(m,n)P_{z}(i,j),
\label{NU2CouloHamil}
\end{align}%
where the Coulomb matrix element is 
\begin{equation}
V_{mnij}^{\pm }=\frac{1}{2}\int \!d^{2}xd^{2}y\,\langle m|\mathbf{x}\rangle
\langle \mathbf{x}|n\rangle V^{\pm }(\mathbf{x}-\mathbf{y})\langle i|\mathbf{%
y}\rangle \langle \mathbf{y}|j\rangle ,
\end{equation}%
and%
\begin{align}
\rho (m,n)=& \sum_{\sigma }c_{\sigma }^{\dag }(m)c_{\sigma }(n),  \notag \\
S_{a}(m,n)=& \frac{1}{2}\sum_{\sigma \tau }c_{\sigma }^{\dag }(m)(\tau _{a}^{%
\text{spin}})_{\sigma \tau }c_{\tau }(n),  \notag \\
P_{a}(m,n)=& \frac{1}{2}\sum_{\sigma \tau }c_{\sigma }^{\dag }(m)(\tau _{a}^{%
\text{ppin}})_{\sigma \tau }c_{\tau }(n).
\end{align}%
They satisfy the W$_{\infty }(4)$ algebra\cite{EzawaX04B}, which is the
SU(4) extension of the W$_{\infty }$ algebra. The Zeeman and pseudo-Zeeman
terms read%
\begin{equation}
H_{\text{ZpZ}}=-\sum_{n}\left[ \Delta _{\text{Z}}S_{z}(n,n)+\Delta _{\text{%
SAS}}P_{x}(n,n)+\Delta _{\text{bias}}P_{z}(n,n)\right] .
\end{equation}%
The total Hamiltonian is $H=H_{\text{C}}^{+}+H_{\text{C}}^{-}+H_{\text{ZpZ}}$%
.

\section{Physical Degrees of Freedom}

\label{SecPhysiDegre}

At $\nu =2$ there are two electrons in one Landau site. We consider a
creation operator of a pair of electrons at site $n$,%
\begin{equation}
G^{\dag }(n)=\frac{1}{2}\sum_{\mu \nu }g_{\mu \nu }c_{\mu }^{\dag }(n)c_{\nu
}^{\dag }(n),
\end{equation}%
where $g_{\mu \nu }$ is an antisymmetric complex matrix, $g_{\mu \nu
}=-g_{\nu \mu }$. A homogeneous state is given by%
\begin{equation}
|\text{g}\rangle =\prod_{n}G^{\dag }(n)|0\rangle ,  \label{NU2EnergF}
\end{equation}%
when $g_{\mu \nu }$ is independent of the site index $n$. The normalization
of the state, $\langle $g$|$g$\rangle =1$, leads to 
\begin{equation}
\text{Tr}\left( gg^{\dag }\right) =2.  \label{NU2GrounNorma}
\end{equation}%
Since $g_{\mu \nu }$ contains $6$ independent complex parameters, there are $%
6$ independent homogeneous states. They span the $\mathbf{6}$-dimensional
irreducible representation of SU(4)$\otimes $SU(4).

It is hard to diagonalize the total Hamiltonian. Since we are interested in
the regime where the SU(4)-invariant Coulomb term $H_{C}^{+}$ dominates all
other interactions, we start with the ground state of the Hamiltonian $%
H_{C}^{+}$,%
\begin{equation}
H_{C}^{+}|\text{g}\rangle =E_{\text{g}}^{+}|\text{g}\rangle .
\label{NU2EnergP}
\end{equation}%
We include the SU(4)-noninvariant terms as small perturbation.

By requiring%
\begin{equation}
\rho (m,n)|\text{g}\rangle =\nu \delta _{mn}|\text{g}\rangle ,
\label{NU2CondiGrounA}
\end{equation}%
the eigenvalue equation (\ref{NU2EnergP}) is satisfied with\cite{EzawaX04B}%
\begin{equation}
E_{\text{g}}^{+}=-\nu \sum_{mn}V_{mnnm}^{+}=-\nu \varepsilon _{X}^{+}N_{\Phi
}  \label{NU2EnergE}
\end{equation}%
and 
\begin{equation}
\varepsilon _{X}^{\pm }=\frac{1}{4}\sqrt{\frac{\pi }{2}}\left[ 1\pm e^{\frac{%
1}{2}(d/\ell )^{2}}\text{erfc}\left( \frac{d}{\sqrt{2}\ell }\right) \right]
E_{\text{C}}^{0}.  \label{ParamPM}
\end{equation}%
Here%
\begin{equation}
E_{\text{C}}^{0}=\frac{e^{2}}{4\pi \varepsilon \ell _{B}}
\end{equation}%
is the Coulomb energy unit. Note that the direct energy part, $\nu
^{2}\sum_{mi}V_{mmii}^{+}$, is cancelled out by the background neutralizing
charge\cite{EzawaX04B}. It is the ground state in the SU(4)-invariant limit
of the system, which is the unperturbed system realized in the limits $%
d\rightarrow 0$, $\Delta _{\text{Z}}\rightarrow 0$, $\Delta _{\text{SAS}%
}\rightarrow 0$ and $\Delta _{\text{bias}}\rightarrow 0$.

We introduce the expectation value of isospin operators, 
\begin{align}
\mathit{S}_{a}=& \langle \text{g}|S_{a}(n,n)|\text{g}\rangle =\frac{1}{2}%
\text{Tr}\left( \tau _{a}^{\text{spin}}gg^{\dag }\right) ,  \notag \\
\mathit{P}_{a}=& \langle \text{g}|P_{a}(n,n)|\text{g}\rangle =\frac{1}{2}%
\text{Tr}\left( \tau _{a}^{\text{ppin}}gg^{\dag }\right) ,  \notag \\
\mathit{R}_{ab}=& \langle \text{g}|R_{ab}(n,n)|\text{g}\rangle =\frac{1}{2}%
\text{Tr}\left( \tau _{a}^{\text{spin}}\tau _{b}^{\text{ppin}}gg^{\dag
}\right) .  \label{NU2StepN}
\end{align}%
They are the total spin per one Landau site, and so on. We pay a special
attention to the imbalance parameter $\sigma _{0}$, which is the normalized
density deference between the two layers, $|\sigma _{0}|\leq 1$. At $\nu =2$
it is defined by 
\begin{equation}
\sigma _{0}\equiv \frac{\rho _{\text{f}}-\rho _{\text{b}}}{\rho _{\text{f}%
}+\rho _{\text{b}}}=\mathit{P}_{z}.
\end{equation}%
The relation (\ref{NU2StepN}) together with the normalization condition (\ref%
{NU2GrounNorma}) yields 
\begin{equation}
gg^{\dag }=\frac{1}{2}\mathbb{I}+\frac{1}{2}\left( \tau _{a}^{\text{spin}}%
\mathit{S}_{a}+\tau _{a}^{\text{ppin}}\mathit{P}_{a}+\tau _{a}^{\text{spin}%
}\tau _{b}^{\text{ppin}}\mathit{R}_{ab}\right) .  \label{NU2gg}
\end{equation}%
Here and hereafter the summation over repeated indices over the spin or the
pseudospin is understood; for instance, $\mathbf{\mathit{S}}^{2}=\mathit{S}%
_{a}^{2}=\sum_{a=xyz}\mathit{S}_{a}\mathit{S}_{a}$.

We study the condition (\ref{NU2CondiGrounA}). It holds trivially for $m=n$.
The condition $\rho (m,n)|$g$\rangle =0$ for $m\neq n$ yields%
\begin{equation}
\sum_{\alpha \beta \mu \nu }g_{\alpha \beta }g_{\mu \nu }c_{\alpha
}^{\dagger }(m)c_{\beta }^{\dagger }(m)c_{\mu }^{\dagger }(m)c_{\nu
}^{\dagger }(n)|0\rangle =0,
\end{equation}%
which in turn leads to%
\begin{equation}
\sum_{\alpha \beta \mu \nu }\epsilon _{\alpha \beta \mu \nu }g_{\alpha \beta
}g_{\mu \nu }=0,  \label{NU2CondiGrounB}
\end{equation}%
where $\epsilon _{\alpha \beta \mu \nu }$ is the totally antisymmetric
tensor. It is equivalent to%
\begin{equation}
\mathbf{\mathit{S}}^{2}+\mathbf{\mathit{P}}^{2}+\mathbf{\mathit{R}}^{2}=1,
\label{NU2CondiA}
\end{equation}%
in terms of physical variables, as we verify in Appendix \ref{AppenGSC}.

It is interesting to note that the magnitude of the SU(4) isospin, $\mathbf{%
\mathit{S}}^{2}+\mathbf{\mathit{P}}^{2}+\mathbf{\mathit{R}}^{2}$, is fixed
by the eigenvalue equation (\ref{NU2EnergP}). Namely it is a dynamical
variable. Indeed, we have%
\begin{equation}
\mathbf{\mathit{S}}^{2}+\mathbf{\mathit{P}}^{2}+\mathbf{\mathit{R}}^{2}=\cos
^{2}2\theta
\end{equation}%
for the state%
\begin{equation}
|\text{g}\rangle =\prod_{n}\left[ \cos \theta \,c_{\text{f}\uparrow
}^{\dagger }(n)c_{\text{f}\downarrow }^{\dagger }(n)+\sin \theta \,c_{\text{b%
}\uparrow }^{\dagger }(n)c_{\text{b}\downarrow }^{\dagger }(n)\right]
|0\rangle .  \label{StateX}
\end{equation}%
In general, we can only derive a kinematical constraint%
\begin{equation}
\mathbf{\mathit{S}}^{2}+\mathbf{\mathit{P}}^{2}+\mathbf{\mathit{R}}%
^{2}\leqslant 1  \label{NU2CondiM}
\end{equation}%
from their definition (\ref{NU2StepN}): See (\ref{NU2CondiL}) in Appendix %
\ref{AppenUV}. This is in sharp contrast to the $\nu =1$ case, where the
magnitude of the SU(4) isospin, $\mathbf{\mathit{S}}^{2}+\mathbf{\mathit{P}}%
^{2}+\mathbf{\mathit{R}}^{2}=3/4$, is fixed kinematically.

Though there are $12$ real parameters in the antisymmetric matrix $g_{\mu
\nu }$, one of them is unphysical for representing the overall phase, and
another is fixed by the normalization condition (\ref{NU2GrounNorma}). Then
one variable is fixed by the condition (\ref{NU2CondiA}) on the ground state 
$|$g$\rangle $. As we shall see soon [see (\ref{NU2StepG})] and prove in
Appendix \ref{AppenGSC}, there exists another unphysical variable which
decouples on the ground state $|$g$\rangle $. Hence the parameter space
characterizing the SU(4)-invariant ground state $|$g$\rangle $ contains 8
real independent variables. They are the 4 complex Goldstone modes
associated with a spontaneous breakdown of the SU(4) symmetry in the
SU(4)-invariant limit. The number of the Goldstone modes agrees with our
previous result\cite{Hasebe02B,EzawaX03B} obtained based on an effective
theory with the use of composite bosons.

We now include the SU(4)-noninvariant interactions as small perturbation.
The first order perturbation is to diagonalize the full Hamiltonian $H$
within this subspace. Equivalently, the ground state is determined by
minimizing the energy 
\begin{equation}
E_{\text{g}}\equiv \mathcal{E}_{\text{g}}N_{\Phi }=\langle \text{g}|H|\text{g%
}\rangle
\end{equation}%
within this parameter space. See Section \ref{SecDiscus} on this point.

It is straightforward to show that%
\begin{equation}
\langle \text{g}|c_{\mu }^{\dag }(m)c_{\nu }(n)|\text{g}\rangle =\delta
_{mn}(gg^{\dag })_{\nu \mu },
\end{equation}%
and%
\begin{align}
& \langle \text{g}|c_{\mu }^{\dag }(m)c_{\sigma }^{\dag }(i)c_{\tau
}(j)c_{\nu }(n)|\text{g}\rangle  \notag \\
=& \left\{ 
\begin{array}{ll}
+\delta _{mn}g_{\mu \sigma }^{\dag }g_{\nu \tau } & \text{for\quad }m=i,j=n
\\ 
+\delta _{mn}\delta _{ij}(gg^{\dag })_{\tau \sigma }(gg^{\dag })_{\nu \mu }
& \text{for\quad }m>i,j<n \\ 
-\delta _{mj}\delta _{in}(gg^{\dag })_{\nu \sigma }(gg^{\dag })_{\tau \mu }
& \text{for\quad }m>i,j>n \\ 
-\delta _{mj}\delta _{in}(gg^{\dag })_{\nu \sigma }(gg^{\dag })_{\tau \mu }
& \text{for\quad }m<i,j<n \\ 
+\delta _{mn}\delta _{ij}(gg^{\dag })_{\tau \sigma }(gg^{\dag })_{\nu \mu }
& \text{for\quad }m<i,j>n%
\end{array}%
\right. .
\end{align}%
Thus%
\begin{align}
\frac{1}{N_{\phi }}\sum & V_{mnij}^{\pm }\langle G|c_{\mu }^{\dag
}(m)c_{\sigma }^{\dag }(i)c_{\tau }(j)c_{\nu }(n)|G\rangle  \notag \\
=& (gg^{\dag })_{\tau \sigma }(gg^{\dag })_{\nu \mu }\varepsilon _{\text{D}%
}^{\pm }-(gg^{\dag })_{\nu \sigma }(gg^{\dag })_{\tau \mu }\varepsilon _{%
\text{X}}^{\pm }.
\end{align}%
By using these formulas the ground-state energy per Landau site is
calculated as%
\begin{align}
\mathcal{E}_{\text{g}}=& \varepsilon _{\text{D}}^{-}\left[ \text{Tr}\left(
\tau _{z}^{\text{ppin}}gg^{\dag }\right) \right] ^{2}-\varepsilon _{\text{X}%
}^{-}\text{Tr}\left( \tau _{z}^{\text{ppin}}gg^{\dag }\tau _{z}^{\text{ppin}%
}gg^{\dag }\right)  \notag \\
-& \varepsilon _{\text{X}}^{+}\text{Tr}\left( gg^{\dag }gg^{\dag }\right) -%
\frac{1}{2}\Delta _{\text{Z}}\text{Tr}\left( \tau _{z}^{\text{spin}}gg^{\dag
}\right)  \notag \\
-& \frac{1}{2}\Delta _{\text{SAS}}\text{Tr}\left( \tau _{x}^{\text{ppin}%
}gg^{\dag }\right) -\frac{1}{2}\Delta _{\text{bias}}\text{Tr}\left( \tau
_{z}^{\text{ppin}}gg^{\dag }\right) ,
\end{align}%
or%
\begin{align}
\mathcal{E}_{\text{g}}=& 4\varepsilon _{\text{D}}^{-}\mathit{P}%
_{z}^{2}-\left( \varepsilon _{\text{X}}^{+}-\varepsilon _{\text{X}%
}^{-}\right) \left( \mathit{S}_{a}^{2}+\mathit{P}_{a}^{2}+\mathit{R}%
_{ab}^{2}\right)  \notag \\
& -2\varepsilon _{\text{X}}^{-}\left( \mathit{S}_{a}^{2}+\mathit{P}_{z}^{2}+%
\mathit{R}_{az}^{2}\right) -\varepsilon _{\text{X}}^{+}-\varepsilon _{\text{X%
}}^{-}  \notag \\
& -\Delta _{\text{Z}}\mathit{S}_{z}-\Delta _{\text{SAS}}\mathit{P}%
_{x}-\Delta _{\text{bias}}\mathit{P}_{z},  \label{NU2EnergA}
\end{align}%
where $\varepsilon _{X}^{\pm }$ is given by (\ref{ParamPM}) and%
\begin{equation}
\varepsilon _{\text{D}}^{-}=\frac{1}{4}\frac{d}{\ell }E_{\text{C}}^{0}.
\end{equation}%
It is necessary to express 15 isospin components, $\mathit{S}_{a}$, $\mathit{%
P}_{a}$ and $\mathit{R}_{ab}$ in terms of independent variables.

We may choose $\mathit{S}_{a}$ and $\mathit{P}_{a}$ as 6 independent
variables. We expect that $\mathit{R}_{ab}$ are written in terms of these 6
variables and 4 extra variables. Indeed, as demonstrated in the Appendix \ref%
{AppenPV}, we obtain%
\begin{align}
\mathit{R}_{ab}=& \frac{\mathbf{\mathit{S}}^{2}\mathit{P}_{a}-(\mathbf{%
\mathit{S}}\mathbf{\mathit{P}})\mathit{S}_{a}}{\mathit{S}\mathit{Q}}\hspace*{%
0.5mm}\frac{\mathbf{\mathit{P}}^{2}\mathit{S}_{b}-(\mathbf{\mathit{S}}%
\mathbf{\mathit{P}})\mathit{P}_{b}}{\mathit{P}\mathit{Q}}\mathit{R}_{PS} 
\notag \\
+& \frac{\mathbf{\mathit{S}}^{2}\mathit{P}_{a}-(\mathbf{\mathit{S}}\mathbf{%
\mathit{P}})\mathit{S}_{a}}{\mathit{S}\mathit{Q}}\frac{\mathit{Q}_{b}}{%
\mathit{Q}}\mathit{R}_{PQ}  \notag \\
+& \frac{\mathit{Q}_{a}}{\mathit{Q}}\frac{\mathbf{\mathit{P}}^{2}\mathit{S}%
_{b}-(\mathbf{\mathit{S}}\mathbf{\mathit{P}})\mathit{P}_{b}}{\mathit{PQ}}%
\mathit{R}_{QS}+\frac{\mathit{Q}_{a}}{\mathit{Q}}\frac{\mathit{Q}_{b}}{%
\mathit{Q}}\mathit{R}_{QQ}  \label{NU2StepF}
\end{align}%
with $\mathit{Q}_{a}\equiv \varepsilon _{abc}\mathit{S}_{b}\mathit{P}_{c},$ $%
\mathit{S}=|\mathbf{\mathit{S}}|$, $\mathit{P}=|\mathbf{\mathit{P}}|$ and $%
\mathit{Q}=|\mathbf{\mathit{Q}}|$. Here, $\mathit{R}_{PS}$, $\mathit{R}_{PQ}$%
, $\mathit{R}_{QS}$ and $\mathit{R}_{QQ}$ are the 4 extra variables.
However, it can be proved that only 3 of them are independent, and they are
parametrized as%
\begin{align}
\mathit{R}_{PS}+i\mathit{R}_{QS}=& e^{i\omega }\left( -i\lambda +\frac{1}{%
\xi }\mathit{S}\right) ,  \notag \\
\mathit{R}_{PQ}+i\mathit{R}_{QQ}=& -ie^{i\omega }\xi \mathit{P}.
\label{NU2StepG}
\end{align}%
Thus $\mathit{R}_{ab}$ are expressed in terms of 9 variables; $\xi $, $%
\lambda $, $\omega $ together with $\mathit{S}_{a}$, $\mathit{P}_{a}$. One
variable is unexpectedly unphysical, about which we explain in Appendix \ref%
{AppenUV}.

We show that $\mathit{S}_{x}=\mathit{S}_{y}=\mathit{P}_{y}=0$ for the ground
state. First of all $\mathit{S}_{a}$ and $\mathit{R}_{az}$ rotate as vectors
under the SU$_{\text{spin}}$(2) transformation. So, if we have any
configuration with nonvanishing $\mathit{S}_{x}$ and $\mathit{S}_{y}$, we
can perform an SU$_{\text{spin}}$(2) rotation to increase $\mathit{S}_{z}$,
without affecting $\mathit{S}_{a}^{2}$ and $\mathit{R}_{az}^{2}$, as far as
possible until $\mathit{S}_{x}=\mathit{S}_{y}=0$. This decreases the energy (%
\ref{NU2EnergA}). Similarly, performing an SU$_{\text{ppin}}$(2) rotation in
the $xy$-pseudoplane, we can lower the energy via the tunneling term by
increasing $\mathit{P}_{x}$, without affecting $\mathit{P}_{z}$ and $\mathit{%
R}_{az}^{2}$, as far as possible until $\mathit{P}_{y}=0$.

Substituting $\mathit{S}_{x}=\mathit{S}_{y}=\mathit{P}_{y}=0$ into $R_{ab}$
we come to%
\begin{align}
\sum_{ab}\mathit{R}_{ab}^{2}=& \frac{\mathit{S}_{z}^{2}}{\xi ^{2}}+\lambda
^{2}+\xi ^{2}\left( \mathit{P}_{x}^{2}+\mathit{P}_{z}^{2}\right) ,  \notag \\
\sum_{a}\mathit{R}_{az}^{2}=& \left( \frac{\mathit{S}_{z}^{2}}{\xi ^{2}}%
+\lambda ^{2}\right) \frac{\mathit{P}_{x}^{2}}{\mathit{P}_{x}^{2}+\mathit{P}%
_{z}^{2}}.  \label{NU2EnergG}
\end{align}%
Hence the energy (\ref{NU2EnergA}) yields%
\begin{align}
\mathcal{E}_{\text{g}}=& \left( \varepsilon _{\text{X}}^{-}-\varepsilon _{%
\text{X}}^{+}\right) \left[ \left( 1+\frac{1}{\xi ^{2}}\right) \mathit{S}%
_{z}^{2}+\left( \mathit{P}_{x}^{2}+\mathit{P}_{z}^{2}\right) \left( 1+\xi
^{2}\right) \right]  \notag \\
& -2\varepsilon _{\text{X}}^{-}\mathit{S}_{z}^{2}+\varepsilon _{\text{cap}}%
\mathit{P}_{z}^{2}-2\varepsilon _{\text{X}}^{-}\frac{\mathit{P}_{x}^{2}}{%
\mathit{P}_{x}^{2}+\mathit{P}_{z}^{2}}\frac{\mathit{S}_{z}^{2}}{\xi ^{2}} 
\notag \\
& -\varepsilon _{\text{X}}^{+}-\varepsilon _{\text{X}}^{-}-\Delta _{\text{Z}}%
\mathit{S}_{z}-\Delta _{\text{SAS}}\mathit{P}_{x}-\Delta _{\text{bias}}%
\mathit{P}_{z}  \notag \\
& -\left( \varepsilon _{\text{X}}^{+}-\varepsilon _{\text{X}%
}^{-}+2\varepsilon _{\text{X}}^{-}\frac{\mathit{P}_{x}^{2}}{\mathit{P}%
_{x}^{2}+\mathit{P}_{z}^{2}}\right) \lambda ^{2},  \label{NU2EnergB}
\end{align}%
where we have set%
\begin{equation}
\varepsilon _{\text{cap}}\equiv 4\varepsilon _{\text{D}}^{-}-2\varepsilon _{%
\text{X}}^{-}.  \label{NU2CapacParam}
\end{equation}%
Here $\varepsilon _{\text{cap}}\mathit{P}_{z}^{2}$ is the capacitance energy
per one Landau site. The capacitance parameter (\ref{NU2CapacParam}) is
different from the $\nu =1$ QH system\cite{EzawaX04B}, where $\varepsilon _{%
\text{cap}}^{\nu =1}\equiv 4\varepsilon _{D}^{-}-4\varepsilon _{X}^{-}$.
Note that the energy formula (\ref{NU2EnergA}) holds as it stands also at $%
\nu =1$. The difference arises because $\sum_{a}\mathit{R}_{az}^{2}=\mathit{P%
}_{z}^{2}$ in (\ref{NU2EnergA}) at $\nu =1$ but it is given by (\ref%
{NU2EnergG}) at $\nu =2$.

If we minimize $\mathcal{E}_{\text{g}}$ within the parameter space of the
SU(4)-invariant ground state, we should impose the condition (\ref{NU2CondiA}%
). However, it is instructive to minimize $\mathcal{E}_{\text{g}}$ without
requiring it. Namely, we only assume the kinematical condition (\ref%
{NU2CondiM}), which reads 
\begin{equation}
\left( 1+\frac{1}{\xi ^{2}}\right) \mathit{S}_{z}^{2}+\left( \mathit{P}%
_{x}^{2}+\mathit{P}_{z}^{2}\right) \left( 1+\xi ^{2}\right) +\lambda
^{2}\leq 1.  \label{NU2CondiB}
\end{equation}%
Nevertheless, the equality is easily seen to hold on the minimum-energy
state. Indeed, let us assume that it is realized with the inequality.
However, this is self-contradictory since we can decrease the energy $%
\mathcal{E}_{\text{g}}$ by increasing $\lambda ^{2}$ in (\ref{NU2EnergB}) as
far as the inequality is obeyed; note that $\varepsilon _{\text{X}%
}^{+}>\varepsilon _{\text{X}}^{-}$. The self-contradiction is resolved only
if the equality holds in (\ref{NU2CondiB}),%
\begin{equation}
\left( 1+\frac{1}{\xi ^{2}}\right) \mathit{S}_{z}^{2}+\left( \mathit{P}%
_{x}^{2}+\mathit{P}_{z}^{2}\right) \left( 1+\xi ^{2}\right) +\lambda ^{2}=1.
\label{NU2CondiBx}
\end{equation}%
Namely, even if we minimize the energy $\mathcal{E}_{\text{g}}$ without
imposing (\ref{NU2CondiA}), we reproduce it. Consequently, the variation
approach reproduces the ground-state condition (\ref{NU2CondiGrounA}) within
the class of test functions (\ref{NU2EnergF}).

We note that $\omega $ disappears both from (\ref{NU2EnergB}) and (\ref%
{NU2CondiB}), and hence it is a zero-energy mode even in the
SU(4)-noninvariant case. We shall see that it is related with the rotational
invariance in the $xy$-plane: See (\ref{NU2StepL}).

We eliminate $\lambda ^{2}$ in (\ref{NU2EnergB}) by using (\ref{NU2CondiBx}),%
\begin{align}
\mathcal{E}_{\text{g}}=& 2\varepsilon _{\text{X}}^{-}\frac{\mathit{P}_{z}^{2}%
}{\mathit{P}_{x}^{2}+\mathit{P}_{z}^{2}}\left( 1-\mathit{S}_{z}^{2}\right)
+2\varepsilon _{\text{X}}^{-}\left( 1+\xi ^{2}\right) \mathit{P}_{x}^{2} 
\notag \\
& +\varepsilon _{\text{cap}}\mathit{P}_{z}^{2}-\Delta _{\text{Z}}\mathit{S}%
_{z}-\Delta _{\text{SAS}}\mathit{P}_{x}-\Delta _{\text{bias}}\mathit{P}_{z} 
\notag \\
& -2\varepsilon _{\text{X}}^{+}-2\varepsilon _{\text{X}}^{-}.
\label{NU2EnergC}
\end{align}%
It is clear that we can decrease the energy by increasing $\mathit{S}_{z}$
without affecting other terms in (\ref{NU2EnergC}). This is achieved at by
decreasing $\lambda ^{2}$ until $\lambda ^{2}=0$ in (\ref{NU2CondiBx}),
which yields%
\begin{equation}
\mathit{S}_{z}^{2}+\xi ^{2}\left( \mathit{P}_{x}^{2}+\mathit{P}%
_{z}^{2}\right) =\frac{\xi ^{2}}{1+\xi ^{2}}.  \label{NU2CondiC}
\end{equation}%
We solve this as 
\begin{align}
\mathit{S}_{z}=& \frac{\xi }{\sqrt{1+\xi ^{2}}}\sqrt{1-\alpha ^{2}},  \notag
\\
\mathit{P}_{x}=& \frac{1}{\sqrt{1+\xi ^{2}}}\alpha \sqrt{1-\beta ^{2}}, 
\notag \\
\mathit{P}_{z}=& \frac{1}{\sqrt{1+\xi ^{2}}}\alpha \beta  \label{NU2StepJ}
\end{align}%
in terms of two parameters $|\alpha |\leqslant 1$ and $|\beta |\leqslant 1$.

\begin{figure}[h]
\includegraphics[width=0.4\textwidth]{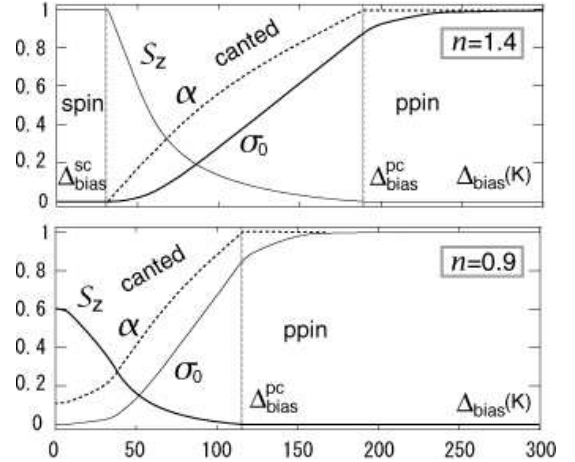}
\caption{The spin component $\mathit{S}_{z}$\ and the imbalance parameter $%
\protect\sigma _{0}\equiv \mathit{P}_{z}$ are illustrated together with $%
\protect\alpha $ as functions of $\Delta _{\text{bias}}$ for typical sample
parameters ($d=23$nm, $\Delta _{\text{SAS}}=6.7$K and $\protect\rho %
_{0}=n\times 10^{11}/$cm$^{2}$). The spin phase ($\protect\alpha =0$), the
canted phase ($0<\protect\alpha <1$) and the ppin phase ($\protect\alpha =1$%
) are realized for $\Delta _{\text{bias}}<\Delta _{\text{bias}}^{\text{sc}}$%
, $\Delta _{\text{bias}}^{\text{sc}}<\Delta _{\text{bias}}<\Delta _{\text{%
bias}}^{\text{pc}}$ and $\Delta _{\text{bias}}^{\text{pc}}<\Delta _{\text{%
bias}}$, respectively. There are 3 phases for $n=1.4$ but only 2 phases for $%
n=0.9$.}
\label{FigNU2Pz}
\end{figure}

Substituting these into (\ref{NU2EnergC}) we obtain 
\begin{align}
\mathcal{E}_{\text{g}}=& 2\varepsilon _{\text{X}}^{-}\alpha ^{2}+\left[
2\varepsilon _{\text{X}}^{-}+4\left( \varepsilon _{\text{D}}^{-}-\varepsilon
_{\text{X}}^{-}\right) \alpha ^{2}\right] \frac{\beta ^{2}}{1+\xi ^{2}} 
\notag \\
-& \frac{\Delta _{\text{Z}}\xi }{\sqrt{1+\xi ^{2}}}\sqrt{1-\alpha ^{2}}-%
\frac{\Delta _{\text{SAS}}}{\sqrt{1+\xi ^{2}}}\alpha \sqrt{1-\beta ^{2}} 
\notag \\
-& \frac{\Delta _{\text{bias}}}{\sqrt{1+\xi ^{2}}}\alpha \beta -2\varepsilon
_{\text{X}}^{+}-2\varepsilon _{\text{X}}^{-}.  \label{NU2EnergD}
\end{align}%
To minimize this with respect to $\alpha $, $\beta $ and $\xi $, we write
down the equations $\partial _{\alpha }E=\partial _{\beta }E=\partial _{\xi
}E=0$. We rearrange them as 
\begin{subequations}
\begin{align}
\Delta _{\text{Z}}^{2}=& \frac{\Delta _{\text{SAS}}^{2}}{1-\beta ^{2}}-\frac{%
4\varepsilon _{\text{X}}^{-}\left( \Delta _{0}^{2}-\beta ^{2}\Delta _{\text{%
SAS}}^{2}\right) }{\Delta _{0}\sqrt{1-\beta ^{2}}},  \label{NU2StepC} \\
\frac{\Delta _{\text{bias}}}{\beta \Delta _{\text{SAS}}}=& \frac{4\left(
\varepsilon _{\text{X}}^{-}+2\alpha ^{2}\left( \varepsilon _{\text{D}%
}^{-}-\varepsilon _{\text{X}}^{-}\right) \right) }{\Delta _{0}}+\frac{1}{%
\sqrt{1-\beta ^{2}}},  \label{NU2StepB} \\
\xi =& \frac{\Delta _{\text{Z}}}{\Delta _{\text{SAS}}}\frac{\sqrt{1-\alpha
^{2}}}{\alpha }\sqrt{1-\beta ^{2}},  \label{NU2StepA}
\end{align}%
where 
\end{subequations}
\begin{equation}
\Delta _{0}\equiv \sqrt{\Delta _{\text{SAS}}^{2}\alpha ^{2}+\Delta _{\text{Z}%
}^{2}\left( 1-\alpha ^{2}\right) \left( 1-\beta ^{2}\right) }.
\end{equation}%
The ground state is determined by these equations. The parameters $\alpha $
and $\beta $ are solved out from (\ref{NU2StepC}) and (\ref{NU2StepB}) in
terms of the sample parameters [FIG.\ref{FigNU2Pz}]. Then, $\xi $ is given
by (\ref{NU2StepA}) in terms of them.

\section{Ground-State Structure}

We discuss the ground-state structure as a function of parameters $\alpha $
and $\beta $. Substituting (\ref{NU2StepA}) into (\ref{NU2StepJ}) we get 
\begin{align}
\mathit{S}_{z}=& \frac{\Delta _{\text{Z}}}{\Delta _{0}}\left( 1-\alpha
^{2}\right) \sqrt{1-\beta ^{2}},  \notag \\
\mathit{P}_{x}=& \frac{\Delta _{\text{SAS}}}{\Delta _{0}}\alpha ^{2}\sqrt{%
1-\beta ^{2}},  \notag \\
\mathit{P}_{z}=& \frac{\Delta _{\text{SAS}}}{\Delta _{0}}\alpha ^{2}\beta .
\label{NU2StepD}
\end{align}%
Recall that $\mathit{S}_{x}=\mathit{S}_{y}=0$ and $\mathit{P}_{y}=0$. Using (%
\ref{NU2StepG}) in (\ref{NU2StepF}), we perform all the necessary
substitutions and obtain 
\begin{align}
\mathit{R}_{xx}+i\mathit{R}_{yx}=& -\frac{\Delta _{\text{SAS}}}{\Delta _{0}}%
\alpha \sqrt{1-\alpha ^{2}}\beta e^{i\omega },  \notag \\
\mathit{R}_{yy}+i\mathit{R}_{xy}=& +\frac{\Delta _{\text{Z}}}{\Delta _{0}}%
\alpha \sqrt{1-\alpha ^{2}}\sqrt{1-\beta ^{2}}e^{i\omega },  \notag \\
\mathit{R}_{xz}+i\mathit{R}_{yz}=& +\frac{\Delta _{\text{SAS}}}{\Delta _{0}}%
\alpha \sqrt{1-\alpha ^{2}}\sqrt{1-\beta ^{2}}e^{i\omega },  \label{NU2StepE}
\end{align}%
while the rest components vanish, $\mathit{R}_{za}=0$. We have expressed all
the isospin components $\mathit{S}_{a}$, $\mathit{P}_{a}$ and $\mathit{R}%
_{ab}$ in terms of three variables $\alpha $, $\beta $ and $\omega $.

Using $2\mathit{S}_{a}^{\text{f}}=\mathit{S}_{a}+\mathit{R}_{az}$ and $2%
\mathit{S}_{a}^{\text{b}}=\mathit{S}_{a}-\mathit{R}_{az}$ we find 
\begin{align}
\mathit{S}_{x}^{\text{f}}& =-\mathit{S}_{x}^{\text{b}}=\frac{1}{2}\frac{%
\Delta _{\text{SAS}}}{\Delta _{0}}\alpha \sqrt{1-\alpha ^{2}}\sqrt{1-\beta
^{2}}\cos \omega ,  \notag \\
\mathit{S}_{y}^{\text{f}}& =-\mathit{S}_{y}^{\text{b}}=\frac{1}{2}\frac{%
\Delta _{\text{SAS}}}{\Delta _{0}}\alpha \sqrt{1-\alpha ^{2}}\sqrt{1-\beta
^{2}}\mathrm{\sin }\omega .  \label{NU2StepL}
\end{align}%
We see that $\omega $ describes the orientation of $\mathit{S}_{a}^{\text{f}%
} $ and $\mathit{S}_{a}^{\text{b}}$ in the $xy$-plane. It is the zero-energy
mode associated with the rotational invariance in the $xy$-plane.

Employing the formulae derived in Appendix \ref{AppenGSC}, the matrix $g$
can be reconstructed from (\ref{NU2StepD}) and (\ref{NU2StepE}). For the
sake of simplicity we take $\omega =\frac{\pi }{4}$ and come to 
\begin{equation}
g=\gamma _{\uparrow }^{\text{s}}g_{\uparrow }^{\text{s}}+\gamma _{0}^{\text{s%
}}g_{0}^{\text{s}}+\gamma _{\downarrow }^{\text{s}}g_{\downarrow }^{\text{s}%
}+\gamma _{\text{f}}^{\text{p}}g_{\text{f}}^{\text{p}}+\gamma _{0}^{\text{p}%
}g_{0}^{\text{p}}+\gamma _{\text{b}}^{\text{p}}g_{\text{b}}^{\text{p}},
\label{NU2StepK}
\end{equation}%
where%
\begin{align}
g_{\uparrow }^{\text{s}}=& \frac{i}{2}\left( 1+\tau _{z}^{\text{spin}%
}\right) \tau _{y}^{\text{ppin}}, & \quad g_{\text{f}}^{\text{p}}=& \frac{i}{%
2}\tau _{y}^{\text{spin}}\left( 1+\tau _{z}^{\text{ppin}}\right) ,  \notag \\
g_{0}^{\text{s}}=& \frac{i}{2}\sqrt{2}\tau _{x}^{\text{spin}}\tau _{y}^{%
\text{ppin}}, & g_{0}^{\text{p}}=& \frac{i}{2}\sqrt{2}\tau _{y}^{\text{spin}%
}\tau _{x}^{\text{ppin}},  \notag \\
g_{\downarrow }^{\text{s}}=& \frac{i}{2}\left( 1-\tau _{z}^{\text{spin}%
}\right) \tau _{y}^{\text{ppin}}, & g_{\text{b}}^{\text{p}}=& \frac{i}{2}%
\tau _{y}^{\text{spin}}\left( 1-\tau _{z}^{\text{ppin}}\right) ,
\end{align}%
and%
\begin{align}
\gamma _{\uparrow }^{\text{s}}=& \frac{1+i}{2\sqrt{2}}\sqrt{1-\alpha ^{2}}%
\left( 1+\frac{\Delta _{\text{Z}}}{\Delta _{0}}\sqrt{1-\beta ^{2}}\right) , 
\notag \\
\gamma _{0}^{\text{s}}=& 0,  \notag \\
\gamma _{\downarrow }^{\text{s}}=& \frac{1-i}{2\sqrt{2}}\sqrt{1-\alpha ^{2}}%
\left( 1-\frac{\Delta _{\text{Z}}}{\Delta _{0}}\sqrt{1-\beta ^{2}}\right) , 
\notag \\
\gamma _{\text{f}}^{\text{p}}=& -\frac{i\alpha }{2}\left( 1+\frac{\Delta _{%
\text{SAS}}}{\Delta _{0}}\beta \right) ,  \notag \\
\gamma _{0}^{\text{p}}=& -\frac{i\alpha }{\sqrt{2}}\frac{\Delta _{\text{SAS}}%
}{\Delta _{0}}\sqrt{1-\beta ^{2}},  \notag \\
\gamma _{\text{b}}^{\text{p}}=& -\frac{i\alpha }{2}\left( 1-\frac{\Delta _{%
\text{SAS}}}{\Delta _{0}}\beta \right) .  \label{NU2StepM}
\end{align}%
Equivalently, the ground state is expressed as%
\begin{align}
|\text{g}(n)\rangle =& \gamma _{\uparrow }^{\text{s}}|\mathit{S}_{\uparrow
}(n)\rangle +\gamma _{0}^{\text{s}}|\mathit{S}_{0}(n)\rangle +\gamma
_{\downarrow }^{\text{s}}|\mathit{S}_{\downarrow }(n)\rangle  \notag \\
& +\gamma _{\text{f}}^{\text{p}}|\mathit{P}_{\text{f}}(n)\rangle +\gamma
_{0}^{\text{p}}|\mathit{P}_{0}(n)\rangle +\gamma _{\text{b}}^{\text{p}}|%
\mathit{P}_{\text{b}}(n)\rangle ,  \label{NU2GrounState}
\end{align}%
where%
\begin{align}
|\mathit{S}_{\uparrow }(n)\rangle =& c_{\text{f}\uparrow }^{\dag }(n)c_{%
\text{b}\uparrow }^{\dag }(n)|0\rangle ,  \notag \\
|\mathit{S}_{0}(n)\rangle =& \frac{c_{\text{f}\uparrow }^{\dag }(n)c_{\text{b%
}\downarrow }^{\dag }(n)+c_{\text{f}\downarrow }^{\dag }(n)c_{\text{b}%
\uparrow }^{\dag }(n)}{\sqrt{2}}|0\rangle ,  \notag \\
|\mathit{S}_{\downarrow }(n)\rangle =& c_{\text{f}\downarrow }^{\dag }(n)c_{%
\text{b}\downarrow }^{\dag }(n)|0\rangle ,  \notag \\
|\mathit{P}_{\text{f}}(n)\rangle =& c_{\text{f}\uparrow }^{\dag }(n)c_{\text{%
f}\downarrow }^{\dag }(n)|0\rangle ,  \notag \\
|\mathit{P}_{0}(n)\rangle =& \frac{c_{\text{f}\uparrow }^{\dag }(n)c_{\text{b%
}\downarrow }^{\dag }(n)-c_{\text{f}\downarrow }^{\dag }c_{\text{b}\uparrow
}^{\dag }(n)}{\sqrt{2}}|0\rangle ,  \notag \\
|\mathit{P}_{\text{b}}(n)\rangle =& c_{\text{b}\uparrow }^{\dag }(n)c_{\text{%
b}\downarrow }^{\dag }(n)|0\rangle .  \label{NU2GrounA}
\end{align}%
This reveals the microscopic structure of the ground state.

All quantities are parametrized by $\alpha $ and $\beta $. When we solve (%
\ref{NU2StepC}) and (\ref{NU2StepB}) for them, we find $\alpha <0$ for $%
\Delta _{\text{bias}}<\Delta _{\text{bias}}^{\text{sc}}$, and $\alpha >1$
for $\Delta _{\text{bias}}>\Delta _{\text{bias}}^{\text{pc}}$ with certain
values of $\Delta _{\text{bias}}^{\text{sc}}$\textbf{\ }and $\Delta _{\text{%
bias}}^{\text{pc}}$: see (\ref{NU2PhaseEx}). Accordingly the ground-state
energy (\ref{NU2EnergD}) is minimized by $\alpha =0$ for $\Delta _{\text{bias%
}}<\Delta _{\text{bias}}^{\text{sc}}$, and $\alpha =1$ for $\Delta _{\text{%
bias}}>\Delta _{\text{bias}}^{\text{pc}}$. Then, the spin component $\mathit{%
S}_{z}$\ and the density imbalance $\mathit{P}_{z}$ are calculated from (\ref%
{NU2StepD}). In FIG.\ref{FigNU2Pz} we illustrate $\mathit{S}_{z}$\ and $%
\sigma _{0}\equiv \mathit{P}_{z}$ together with $\alpha $ as functions of $%
\Delta _{\text{bias}}$ for typical sample parameters; $d=23$nm, $\Delta _{%
\text{SAS}}=6.7$K and $\rho _{0}=n\times 10^{11}/$cm$^{2}$ with $n=1.4$ and $%
=0.9$.

First, when $\alpha =0$, it follows that $\mathit{S}_{z}=1$ and $\mathit{P}%
_{z}=0$ since $\Delta _{0}=\Delta _{\text{Z}}\sqrt{1-\beta ^{2}}$. Note that 
$\beta $ disappears from all formulas in (\ref{NU2StepM}). The spin phase is
characterized by the fact that the isospin is fully polarized into the spin
direction with $\mathit{S}_{z}=1$ and all others being zero. The spins in
both layers point to the positive $z$-axis due to the Zeeman effect.
Substituting $\alpha =0$ and $\Delta _{0}\equiv \Delta _{\text{Z}}\sqrt{%
1-\beta ^{2}}$ into (\ref{NU2StepM}) we find the ground state to be%
\begin{equation}
|\text{g}_{\text{spin}}\rangle =\prod_{n}c_{\text{f}\uparrow }^{\dag }(n)c_{%
\text{b}\uparrow }^{\dag }(n)|0\rangle .  \label{NU2GrounSpin}
\end{equation}%
It is interesting to notice that, even if the bias voltage is applied, no
charge transfer occurs between the two layers ($\sigma _{0}=0$) as far as $%
\Delta _{\text{bias}}<\Delta _{\text{bias}}^{\text{sc}}$, where the system
is in the spin phase [FIG.\ref{FigNU2Pz}].

Second, when $\alpha =1$, it follows that $\mathit{S}_{z}=0$ and $\mathit{P}%
_{z}\neq 0$ (actually $\mathit{P}_{x}^{2}+\mathit{P}_{z}^{2}=1$). The ppin
phase is characterized by the fact that the isospin is fully polarized into
the pseudospin direction with%
\begin{equation}
\mathit{P}_{x}=\sqrt{1-\beta ^{2}},\qquad \mathit{P}_{z}=\beta ,
\end{equation}%
and all others being zero. Because $\mathit{P}_{z}$ represents the density
difference between the two layers, $\beta $ is identified with the imbalance
parameter $\sigma _{0}$. Substituting $\alpha =1$ into (\ref{NU2StepM}) we
obtain the ground state as%
\begin{align}
|\text{g}_{\text{ppin}}\rangle =& \prod_{n}\left\{ \frac{1}{2}\left( \sqrt{%
1+\sigma _{0}}c_{\text{f}\uparrow }^{\dag }(n)+\sqrt{1-\sigma _{0}}c_{\text{b%
}\uparrow }^{\dag }(n)\right) \right.  \notag \\
& \times \left. \left( \sqrt{1+\sigma _{0}}c_{\text{f}\downarrow }^{\dag
}(n)+\sqrt{1-\sigma _{0}}c_{\text{b}\downarrow }^{\dag }(n)\right) \right\}
|0\rangle .  \label{NU2GrounPpin}
\end{align}%
It is found that all electrons are in the front layer when $\sigma _{0}=1$,
and in the back layer when $\sigma _{0}=-1$.

For intermediate values of $\alpha $ ($0<\alpha <1$) none of the spin and
pseudospin vanish, where we may control the density imbalance by applying a
bias voltage as in the ppin phase. The ground state $|$g$_{\text{cant}%
}\rangle $ is given by (\ref{NU2GrounState}) with (\ref{NU2StepM}). All
states except $|\mathit{S}_{0}(n)\rangle $ contribute to form the ground
state. This is so even in the balanced configuration ($\beta =0$). It
follows from (\ref{NU2StepD}) and (\ref{NU2StepL}) that, as the system goes
away from the spin phase, the spin begin to cant and make antiferromagnetic
correlations between the two layers. Hence, it is called the canted
antiferromagnetic phase\cite{Zheng97L}.

We conclude that there are three phases in general; the spin phase, the
canted phase and the ppin phase. They are as characterized by%
\begin{equation}
\begin{tabular}{|l|l|l|}
\hline
spin & canted & ppin \\ \hline
$\mathbf{\mathit{S}}^{2}=1$ & $\mathbf{\mathit{S}}^{2}\neq 0$ & $\mathbf{%
\mathit{S}}^{2}=0$ \\ \hline
$\mathbf{\mathit{P}}^{2}=0$ & $\mathbf{\mathit{P}}^{2}\neq 0$ & $\mathbf{%
\mathit{P}}^{2}=1$ \\ \hline
\end{tabular}%
.
\end{equation}%
The order parameters are $\mathbf{\mathit{S}}^{2}$ and $\mathbf{\mathit{P}}%
^{2}$.

\begin{figure}[h]
\includegraphics[width=0.45\textwidth]{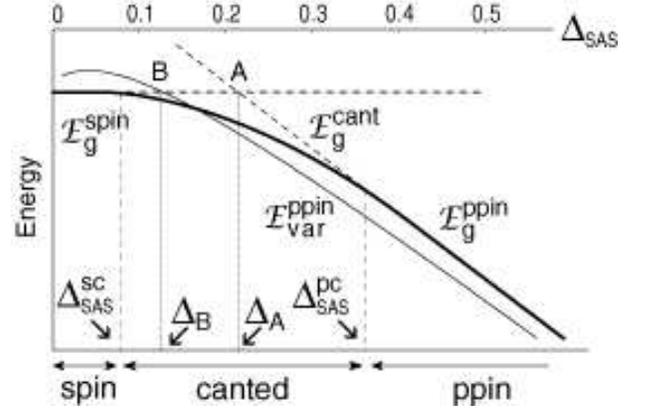}
\caption{The ground-state energies $\mathcal{E}_{\text{g}}^{\text{spin}}$, $%
\mathcal{E}_{\text{g}}^{\text{ppin}}$ and $\mathcal{E}_{\text{g}}^{\text{cant%
}}$ are given in the spin, ppin and canted phases in the balanced
configuration by taking $d=0.56\ell _{B}$. The horizontal axis is the
tunneling gap $\Delta _{\text{SAS}}$ in the Coulomb unit $E_{\text{C}%
}^{0}=e^{2}/(4\protect\pi \protect\varepsilon \ell _{B})$. A phase
transition occurs along the heavy curve continuously from the spin phase to
the canted phase, and then to the ppin phase, as $\Delta _{\text{SAS}}$
increases. The phase transition points $\Delta _{\text{SAS}}^{\text{sc}}$
and $\Delta _{\text{SAS}}^{\text{pc}}$ are given by (\protect\ref{NU2PhaseA}%
) and (\protect\ref{NU2PhaseB}) in the first order of perturbation. The
energy $\mathcal{E}_{\text{g}}^{\text{spin}}$ is not modified but $\mathcal{E%
}_{\text{g}}^{\text{ppin}}$ is modified into $\mathcal{E}_{\text{var}}^{%
\text{ppin}}$ by higher order quantum corrections. The point A stands for
the would-be crossing of the spin and ppin energy levels in the first order
of perturbation, with $\Delta _{A}$ given by (\protect\ref{WouldBeSP}). The
point B stands for the one when higher order quantum corrections are taken
into account, with $\Delta _{B}$ given by (\protect\ref{WouldBeSPx}),
suggesting that the canted phase is considerably shrunk.}
\label{FigFanBL2SCP}
\end{figure}

Let us study the system with no bias voltage in more detail. The spin and
pseudospin are 
\begin{align}
\mathit{S}_{z}=& \frac{\Delta _{\text{Z}}}{\sqrt{\Delta _{\text{SAS}%
}^{2}\alpha ^{2}+\Delta _{\text{Z}}^{2}\left( 1-\alpha ^{2}\right) }}\left(
1-\alpha ^{2}\right) ,  \notag \\
\mathit{P}_{x}=& \frac{\Delta _{\text{SAS}}}{\sqrt{\Delta _{\text{SAS}%
}^{2}\alpha ^{2}+\Delta _{\text{Z}}^{2}\left( 1-\alpha ^{2}\right) }}\alpha
^{2},
\end{align}%
where $\alpha ^{2}$ is easily obtained from (\ref{NU2StepC}) with $\beta =0$%
, 
\begin{equation}
\alpha ^{2}=\frac{\Delta _{\text{SAS}}^{2}-\Delta _{\text{Z}}^{2}}{%
4(2\varepsilon _{\text{X}}^{-})^{2}}-\frac{\Delta _{\text{Z}}^{2}}{\Delta _{%
\text{SAS}}^{2}-\Delta _{\text{Z}}^{2}}.  \label{CanteAlpha}
\end{equation}%
The ground-state energy in each phase is given by%
\begin{align}
\mathcal{E}_{\text{g}}^{\text{spin}}=& -2\varepsilon _{\text{X}%
}^{+}-2\varepsilon _{\text{X}}^{-}-\Delta _{\text{Z}},  \notag \\
\mathcal{E}_{\text{g}}^{\text{cant}}=& -2\varepsilon _{\text{X}}^{+}-\frac{%
\Delta _{\text{SAS}}^{2}}{8\varepsilon _{\text{X}}^{-}}+\frac{\Delta _{\text{%
Z}}^{2}}{8\varepsilon _{\text{X}}^{-}}-\frac{2\varepsilon _{\text{X}%
}^{-}\Delta _{\text{SAS}}^{2}}{\Delta _{\text{SAS}}^{2}-\Delta _{\text{Z}%
}^{2}},  \notag \\
\mathcal{E}_{\text{g}}^{\text{ppin}}=& -2\varepsilon _{\text{X}}^{+}-\Delta
_{\text{SAS}}.  \label{GrounEnergSCP}
\end{align}%
We have depicted them as a function of $\Delta _{\text{SAS}}$ in FIG.\ref%
{FigFanBL2SCP}.

If the canted phase were ignored, the two levels $\mathcal{E}_{\text{g}}^{%
\text{spin}}$ and $\mathcal{E}_{\text{g}}^{\text{ppin}}$ would cross at%
\begin{equation}
\Delta _{\text{SAS}}^{\text{sp}}=\Delta _{\text{Z}}+2\varepsilon _{\text{X}%
}^{-},  \label{WouldBeSP}
\end{equation}%
and a transition would occur suddenly from the spin phase ($\mathbf{\mathit{S%
}}^{2}=1$, $\mathbf{\mathit{P}}^{2}=0$) to the ppin phase ($\mathbf{\mathit{S%
}}^{2}=0$, $\mathbf{\mathit{P}}^{2}=1$). Actually, a mixing of states occurs
and lowers the ground state energy. As a result the level crossing turns
into an level anticrossing [FIG.\ref{FigFanBL2SCP}], and the canted state
emerges between the spin and ppin phases. As $\Delta _{\text{SAS}}$
increases, the phase transition occurs continuously from the spin phase to
the canted phase, and then to the ppin phase. The phase transition points are%
\begin{equation}
\Delta _{\text{SAS}}^{\text{sc}}=\sqrt{\Delta _{\text{Z}}^{2}+4\varepsilon _{%
\text{X}}^{-}\Delta _{\text{Z}}}  \label{NU2PhaseA}
\end{equation}%
from $\mathcal{E}_{g}^{\text{spin}}=\mathcal{E}_{g}^{\text{cant}}$, and%
\begin{equation}
\Delta _{\text{SAS}}^{\text{pc}}=2\varepsilon _{\text{X}}^{-}+\sqrt{\Delta _{%
\text{Z}}^{2}+(2\varepsilon _{\text{X}}^{-})^{2}}  \label{NU2PhaseB}
\end{equation}%
from $\mathcal{E}_{g}^{\text{ppin}}=\mathcal{E}_{g}^{\text{cant}}$. These
two formulas agree with the variational result due to MacDonald et al.\cite%
{MacDonald99B}, where the variational state has been chosen from a subset of
the full set (\ref{NU2EnergF}) satisfying (\ref{NU2CondiGrounA}).

\section{Phase Diagrams}

The phase diagram can be studied based on analytic formulas given in the
previous section. We first present the phase diagram in the $\Delta _{\text{%
SAS}}$-$\Delta _{\text{Z}}$ plane for typical values of $\Delta _{\text{bias}%
}$ to compare our result with the standard ones\cite{Brey99L,MacDonald99B}.
We search for the boundaries separating the canted phase from the spin and
ppin phases in the system. These are extracted from (\ref{NU2StepC}) and (%
\ref{NU2StepB}). The merit of our formalism is that analytic expressions are
available to determine the phase boundaries even for imbalanced
configurations.

Along the spin-canted boundary we have $\alpha =0$. Substituting this value
into (\ref{NU2StepB}) we get%
\begin{equation}
\frac{\beta }{\sqrt{1-\beta ^{2}}}=\frac{\Delta _{\text{bias}}}{\Delta _{%
\text{SAS}}}\frac{\Delta _{\text{Z}}}{\Delta _{\text{Z}}+4\varepsilon _{%
\text{X}}^{-}}.  \label{NU2PhaseC}
\end{equation}%
We solve this for $\beta $ and substitute it into (\ref{NU2StepC}) to get%
\begin{equation}
\Delta _{\text{SAS}}^{2}=\Delta _{\text{Z}}^{2}+4\varepsilon _{\text{X}%
}^{-}\Delta _{\text{Z}}-\frac{\Delta _{\text{Z}}\Delta _{\text{bias}}^{2}}{%
\Delta _{\text{Z}}+4\varepsilon _{\text{X}}^{-}}.  \label{NU2PhaseE}
\end{equation}%
This determines the spin-canted boundary.

\begin{figure}[h]
\includegraphics[width=0.48\textwidth]{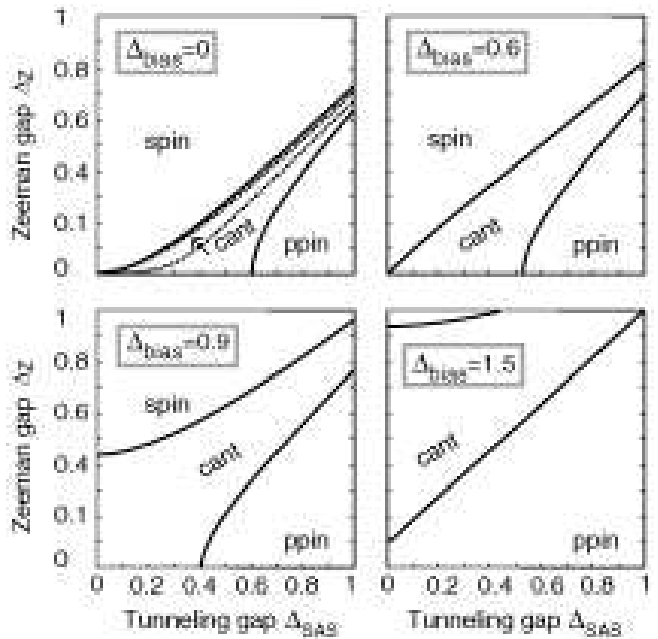}
\caption{The phase diagram is given in the $\Delta _{\text{SAS}}$-$\Delta _{%
\text{Z}}$ plane by changing the bias energy $\Delta _{\text{bias}}$. Here
we have set $d=\ell _{B}$, and taken the Coulomb energy unit $E_{\text{C}%
}^{0}$ for the tunneling gap (horizontal axis) and the Zeeman gap (vertical
axis). It is observed that both the canted and ppin phases are stabilarized
even for $\Delta _{\text{SAS}}=0$ by applying the bias voltage. The dotted
curves in the top left panel represent the exact diagonalization result due
to Schliemann et al.\protect\cite{Schliemann00L} for the 12 electron system.
It is observed that the ppin-canted phase boundary is modified considerably
by higher oder quantum corrections.}
\label{FigPD-ZSAS}
\end{figure}

Along the ppin-canted boundary we have $\alpha =1$. Substituting this value
into (\ref{NU2StepC}) and (\ref{NU2StepB}) we get%
\begin{equation}
\Delta _{\text{SAS}}=\sqrt{1-\beta ^{2}}\left[ \frac{\Delta _{\text{bias}}}{%
\beta }-2\varepsilon _{\text{cap}}\right]  \label{NU2PhaseDx}
\end{equation}%
and%
\begin{equation}
\Delta _{\text{Z}}^{2}=\left( \frac{\Delta _{\text{bias}}}{\beta }%
-2\varepsilon _{\text{cap}}\right) \left( \frac{\Delta _{\text{bias}}}{\beta 
}-8\varepsilon _{\text{D}}^{-}+4\beta ^{2}\varepsilon _{\text{X}}^{-}\right)
\label{NU2PhaseDy}
\end{equation}%
after some manipulation. These two equations give a parametric
representation of the ppin-canted boundary in terms of $\beta $.

\begin{figure}[h]
\includegraphics[width=0.48\textwidth]{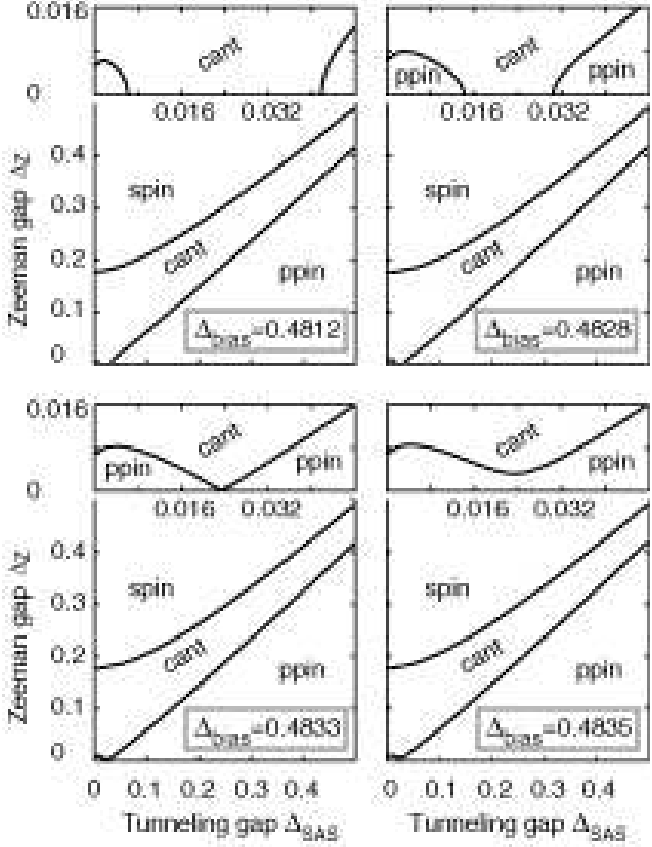}
\caption{The phase diagram is given in the $\Delta _{\text{SAS}}$-$\Delta _{%
\text{Z}}$ plane by changing the bias energy $\Delta _{\text{bias}}$. Here
we have set $d=0.4\ell _{B}$, and taken the Coulomb energy unit $E_{\text{C}%
}^{0}$ for the tunneling gap (horizontal axis) and the Zeeman gap (vertical
axis). A small ppin region appears in the vicinity of the origin for $%
d\lesssim 0.75\ell _{B}$. This region is isolated from the basic ppin region
by the canted phase. This is shown in the top left panel. The upper part of
each panel is the enlarged image of the vicinity of origin where the
isolated ppin region appears. }
\label{FigPD-ZSAS2}
\end{figure}
In drawing the phase diagram in the $\Delta _{\text{SAS}}$-$\Delta _{\text{Z}%
}$ plane we need to fix the layer separation $d$. Following the standard
literature\cite{Brey99L,MacDonald99B}, we have examined the case $d=\ell
_{B} $ and presented the phase diagram in FIG.\ref{FigPD-ZSAS}. Our results
agree qualitatively with results obtained numerically by Brey et al.\cite%
{Brey99L} and MacDonald et al.\cite{MacDonald99B} for imbalanced
configurations.

Our analytic formulas reveal some new features not reported in literature.
We have found some peculiar behaviors for the ppin-canted boundary as in FIG.%
\ref{FigPD-ZSAS2}, which occurs for $2\varepsilon _{\text{D}%
}^{-}<3\varepsilon _{\text{X}}^{-}$, or $d\lesssim 0.75\ell _{B}$. When the
bias voltage is increased, a small ppin region appears in the vicinity of
the origin at 
\begin{equation}
\Delta _{\text{B}}^{(1)}=2\varepsilon _{\text{cap}}.  \label{NU2StepH}
\end{equation}%
This region is isolated from the basic ppin region by the canted phase. By
increasing $\Delta _{\text{bias}}$, the isolated and basic ppin regions come
closer, and they merge at 
\begin{equation}
\Delta _{\text{B}}^{(2)}=8\left[ \frac{2\varepsilon _{\text{D}}^{-}}{%
3\varepsilon _{\text{X}}^{-}}\right] ^{\frac{3}{2}}\varepsilon _{\text{X}%
}^{-}.  \label{NU2StepI}
\end{equation}%
Then the isolated region disappears.

These phase diagrams are, however, not so useful to analyze experimental
data, since we need many samples with different $d$ to realize, e.g., $%
d=\ell _{B}$, but this is impossible. It is more interesting to see
experimentally\cite{Sawada98L} how the phase transition occurs by
controlling the total density $\rho _{0}$ and also the imbalance parameter $%
\sigma _{0}$ in a single sample with fixed values of $d$ and $\Delta _{\text{%
SAS}}$. Here we wish to describe new aspects of the phase diagram from this
point of view.

\begin{figure}[h]
\includegraphics[width=0.36\textwidth]{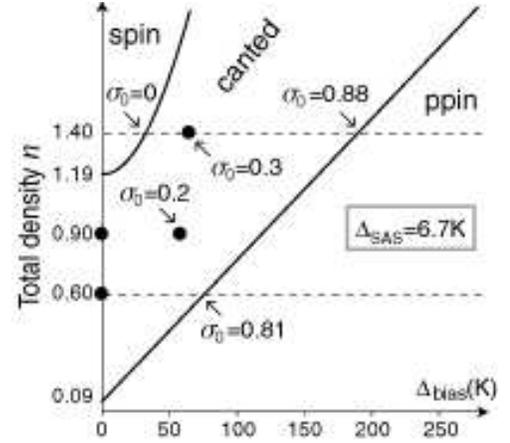}
\caption{The phase diagram is given in the $\Delta _{\text{bias}}$-$\protect%
\rho _{0}$ plane for the sample with $d=23$nm, where $\protect\rho %
_{0}=n\times 10^{11}/$cm$^{-2}$. The two curves stand for the spin-canted
phase boundary and the ppin-canted phase boundary in the first order of
perturbation. The four large points are experimental data indicating phase
transition points taken from Sawada et al.\protect\cite{Sawada98L}. }
\label{FigNU2nBias67}
\end{figure}

\begin{figure}[h]
\includegraphics[width=0.36\textwidth]{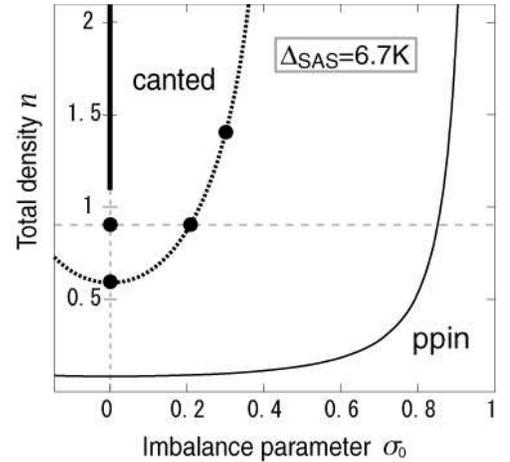}
\caption{The phase diagram is given in the $\protect\sigma _{0}$-$\protect%
\rho _{0}$ plane, where the sample parameters are the same as in FIG. 
\protect\ref{FigNU2nBias67}. The spin phase is realized only in the balance
configuration ($\protect\sigma _{0}=0$) along the heavy line. The solid
curve represents the ppin-canted phase boundary in the first order of
perturbation. The four large points are experimental data indicating phase
transition points taken from Sawada et al.\protect\cite{Sawada98L}. The
dotted curve is a speculated ppin-canted phase boudary. The dotted
horizontal line at $n=0.9$ corresponds to the plateau-width curve indexed by 
$n=0.9$ in FIG.\protect\ref{FigSawadaPRL}(b).}
\label{FigNU2nPz67}
\end{figure}

We start with the spin phase in the balanced configuration with $\Delta _{%
\text{bias}}=0$. (The spin phase realizes provided the electron density is
more than a certain critical value.) When the bias voltage is applied ($%
\Delta _{\text{bias}}\neq 0$) the parameter $\beta $ becomes nonzero
according to (\ref{NU2PhaseC}). However, no charge imbalance is induced
between the two layers ($\sigma _{0}=0$) as far as $\alpha =0$. A charge
imbalance occurs only above a certain critical value $\Delta _{\text{bias}}^{%
\text{sc}}$\ of the bias voltage as in FIG.\ref{FigNU2Pz}, which is given by
solving (\ref{NU2PhaseE}) as 
\begin{equation}
\left( \Delta _{\text{bias}}^{\text{sc}}\right) ^{2}=(\Delta _{\text{Z}%
}+4\varepsilon _{\text{X}}^{-})^{2}-\left( 1+4\frac{\varepsilon _{\text{X}%
}^{-}}{\Delta _{\text{Z}}}\right) \Delta _{\text{SAS}}^{2}.
\label{NU2PhaseEx}
\end{equation}%
This gives the spin-canted phase boundary in the $\Delta _{\text{bias}}$-$%
\rho _{0}$ plane [FIG.\ref{FigNU2nBias67}]. For $\Delta _{\text{bias}%
}>\Delta _{\text{bias}}^{\text{sc}}$ the system is driven into the canted
phase with $\alpha \neq 0$. As the bias voltage increases above a second
critical value $\Delta _{\text{bias}}^{\text{pc}}$, the system turns into
the ppin phase with $\alpha =1$ as in FIG.\ref{FigNU2Pz}. The critical value 
$\Delta _{\text{bias}}^{\text{pc}}$ is obtained by eliminating $\beta $ in (%
\ref{NU2PhaseDx}) and (\ref{NU2PhaseDy}). This gives the ppin-canted phase
boundary in the $\Delta _{\text{bias}}$-$\rho _{0}$ plane [FIG.\ref%
{FigNU2nBias67}]. We remark that, as the electron density decreases, the
critical point $\Delta _{\text{bias}}^{\text{sc}}$ decreases and eventually
becomes zero so that the spin phase disappears at all: The critical density
is $\rho _{0}=1.19\times 10^{11}$/cm$^{-2}$ in the case of FIG.\ref%
{FigNU2nBias67}.

To compare the experimental data\cite{Sawada98L,Sawada99B} it is more
convenient to present the phase diagram in the $\sigma _{0}$-$\rho _{0}$
plane [FIG.\ref{FigNU2nPz67}], which is constructed by using the relation
between the bias voltage and the imbalance parameter implied by (\ref%
{NU2PhaseDx}) and (\ref{NU2PhaseDy}).

\begin{figure}[h]
\includegraphics[width=0.4\textwidth]{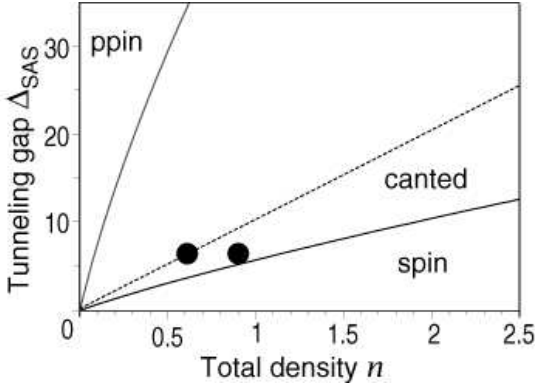}
\caption{The phase diagram is given in the $\protect\rho _{0}$-$\Delta _{%
\text{SAS}}$ plane at the balance point, where the sample parameters are the
same as in FIG. \protect\ref{FigNU2nBias67}. The horizontal axis denotes the
total density $\protect\rho _{0}=n\times 10^{11}/$cm$^{2}$, while the
vertical axis denotes the tunneling gap $\Delta _{\text{SAS}}$ in Kelvin.
The dotted curve is a speculated ppin-canted phase boundary. The two large
points are experimental data indicating phase transition points taken from
Sawada et al.\protect\cite{Sawada98L}. }
\label{FigNU2DsasN}
\end{figure}

It is also useful to study the phase diagram in the $\rho _{0}$-$\Delta _{%
\text{SAS}}$ plane, since it is not difficult to prepare samples with
different $\Delta _{\text{SAS}}$ with all other parameters unchanged [FIG.%
\ref{FigNU2DsasN}]. This is constructed from (\ref{NU2PhaseA}) and (\ref%
{NU2PhaseB}).

\section{Higher Order Corrections}

We have explored the ground-state structure in the first order of
perturbation theory. Our Hamiltonian consists of the SU(4)-invariant term $%
H_{\text{C}}^{+}$ and the SU(4)-noninvariant term $H_{\text{non}}=H_{\text{C}%
}^{-}+H_{\text{ZpZ}}$. We are interested in the regime where $H_{C}^{+}\gg
H_{\text{non}}$. Hence we have diagonalized $H_{\text{C}}^{+}$ as in (\ref%
{NU2EnergP}) and treat $H_{\text{non}}$ perturbatively. The eigenstates are
degenerate with respect to $H_{\text{C}}^{+}$, consisting of 6 states at
each Landau site. The degeneracy is removed by the SU(4)-noninvariant terms $%
H_{\text{non}}$. We may construct $6$ states $|$g$_{i}\rangle $ satisfying%
\begin{equation}
\langle \text{g}_{j}|\text{g}_{i}\rangle =\delta _{ij},\qquad \langle \text{g%
}_{j}|H|\text{g}_{i}\rangle =E_{i}\delta _{ij}  \label{IMFTa}
\end{equation}%
as follows. First we determine $|$g$_{0}\rangle $ by minimizing $\langle $g$%
|H|$g$\rangle $ with the use of a general state (\ref{NU2GrounState}). We
then construct a Fock space made of $5$ states that are orthogonal to $|$g$%
_{0}\rangle $. Within this space we can minimize $\langle $g$|H|$g$\rangle $%
, which determines the lowest energy state $|$g$_{1}\rangle $. In this way
we may construct $6$ states $|$g$_{0}\rangle $, $\cdots $, $|$g$_{5}\rangle $
with $E_{0}\leq E_{1}\leq \cdots \leq E_{5}$, which satisfy (\ref{IMFTa}).
The full Hamiltonian is diagonalized within the first order perturbation
theory.

We have constructed the phase boundaries between the spin phase, the canted
phase and the ppin phase. Here we recall the exact-diagonalization result%
\cite{Schliemann00L} of a few-electron system showing that the boundary
between the spin and canted phases is practically unmodified but the
boundary between the canted and ppin phases is considerably modified from
the mean-field result [FIG.\ref{FigPD-ZSAS}]. In order to explain this we
discuss effects due to higher order perturbations.

We start with the spin phase. It is important that the fully spin polarized
state (\ref{NU2GrounSpin}) is an exact eigenstate of the total Hamiltonian $%
H $. This follows from%
\begin{equation}
P_{z}(i,j)|\text{g}_{\text{spin}}\rangle =P_{x}(i,j)|\text{g}_{\text{spin}%
}\rangle =0
\end{equation}%
and%
\begin{equation}
S_{z}(i,i)|\text{g}_{\text{spin}}\rangle =1.
\end{equation}%
We obtain $H|$g$_{\text{spin}}\rangle =\mathcal{E}_{\text{g}}^{\text{spin}}|$%
g$_{\text{spin}}\rangle $ with 
\begin{equation}
\mathcal{E}_{\text{g}}^{\text{spin}}=-\left( 2\varepsilon _{\text{X}%
}^{+}+2\varepsilon _{\text{X}}^{-}+\Delta _{\text{Z}}\right) .
\label{NU2gEnergS}
\end{equation}%
The eigenvalue agrees with the mean-field value in (\ref{GrounEnergSCP}).
Hence the mean-field equations (\ref{NU2StepC}) $\sim $ (\ref{NU2StepA}) are
exact ones in the spin phase, implying that the spin-canted phase boundary
is not affected by higher order perturbations.

We proceed to discuss the ppin phase. For simplicity we consider the
balanced point, where%
\begin{equation}
S_{z}(i,i)|\text{g}_{\text{ppin}}\rangle =1,\quad P_{z}(i,i)|\text{g}_{\text{%
ppin}}\rangle =0.
\end{equation}%
However, since%
\begin{equation}
P_{z}(i,j)|\text{g}_{\text{ppin}}\rangle \neq 0\quad \text{for}\quad i\neq j,
\end{equation}%
though $\langle g_{\text{ppin}}|P_{z}(i,j)|$g$_{\text{ppin}}\rangle =0$, the
state $|$g$_{\text{ppin}}\rangle $ is not an eigenstate of the capacitance
term $H_{\text{C}}^{-}$. Hence the fully pseudospin polarized state (\ref%
{NU2GrounPpin}) is not an eigenstate of the total Hamiltonian $H$ even in
the absence of the tunneling interaction ($\Delta _{\text{SAS}}=0$).

It is practically impossible to carry out the second order perturbation
since almost no eigenstates of the unperturbed Hamiltonian $H_{\text{C}}^{+}$
are known. We make a variational analysis. We decompose the total
Hamiltonian $H$ into two pieces, $H_{0}$ and $H_{1}$, where $H_{0}$ is the
maximal part of $H$ one of whose eigenstates is $|$g$_{\text{ppin}}\rangle $%
. We do not give their explicit forms here since they are very complicated.
We obtain 
\begin{equation}
H_{0}|g_{\text{ppin}}\rangle =\mathcal{E}_{\text{g}}^{\text{ppin}}|g_{\text{%
ppin}}\rangle
\end{equation}%
with%
\begin{equation}
\mathcal{E}_{\text{g}}^{\text{ppin}}=-2\varepsilon _{\text{X}%
}^{+}-\varepsilon _{\text{cap}}\sigma _{0}^{2}-\frac{\Delta _{\text{SAS}}}{%
\sqrt{1-\sigma _{0}^{2}}}.
\end{equation}%
This agrees with the first-order perturbation result (\ref{GrounEnergSCP})
at $\sigma _{0}=0$. Hence the variational analysis surely presents a higher
order correction. According to the standard procedure, we minimize the total
energy with the variational state 
\begin{equation}
|\text{g}_{\text{ppin}}^{\text{var}}\rangle =(1+\lambda H_{1})|\text{g}_{%
\text{ppin}}\rangle ,
\end{equation}%
where $\lambda $ is the variational parameter. Then the energy reads%
\begin{equation}
\mathcal{E}_{\text{var}}^{\text{ppin}}=\mathcal{E}_{\text{g}}^{\text{ppin}}+%
\frac{|\langle \text{g}_{\text{ppin}}|H_{1}^{2}|\text{g}_{\text{ppin}%
}\rangle |^{2}}{\mathcal{E}_{\text{g}}^{\text{ppin}}\langle \text{g}_{\text{%
ppin}}|H_{1}^{2}|\text{g}_{\text{ppin}}\rangle -\langle \text{g}_{\text{ppin}%
}|H_{1}H_{0}H_{1}|\text{g}_{\text{ppin}}\rangle }.
\end{equation}%
The calculation is straightforward though quite tedious. To perform various
integrals explicitly we approximate (\ref{CouloPM}) as%
\begin{eqnarray}
V^{+}(\mathbf{q}) &=&\frac{1}{|\mathbf{q}|}\exp \left( -\frac{1}{2}|\mathbf{q%
}|^{2}\right) E_{\text{C}}^{0},  \notag \\
V^{-}(\mathbf{q}) &=&\frac{d}{2}\exp \left( -\frac{1}{2}|\mathbf{q}%
|^{2}\right) E_{\text{C}}^{0}
\end{eqnarray}%
with $E_{\text{C}}^{0}=e^{2}/(4\pi \varepsilon \ell _{B})$, which is valid
for $d/\ell _{B}\ll 1$. The result is given by%
\begin{equation}
\mathcal{E}_{\text{var}}^{\text{ppin}}=\mathcal{E}_{\text{g}}^{\text{ppin}}-%
\frac{d^{2}}{16\ell _{B}^{2}}\frac{(1-\sigma _{0}^{2})^{2}E_{\text{C}}^{0}}{%
\sqrt{\pi }\left( \frac{1}{2}+\frac{1}{\sqrt{2}}-\frac{2}{\sqrt{3}}\right) +%
\frac{\Delta _{\text{SAS}}/E_{\text{C}}^{0}}{\sqrt{1-\sigma _{0}^{2}}}}.
\label{NU2gEnergP}
\end{equation}%
The correction is larger for larger layer separation $d$, and it is largest
at the balanced point ($\sigma _{0}=0$).

It is not easy to derive higher order corrections for the canted phase,
since $|$g$_{\text{cant}}\rangle $ is not an eigenstate of any simple
Hamiltonian and furthermore it is a complicated state involving 5 states as
in (\ref{NU2GrounState}). Though we are unable to determine the improved
ppin-canted phase boundary, we may make some arguments how it is modified.
For simplicity we study the balanced point. We have shown that the would-be
phase transition point is $\Delta _{\text{SAS}}^{\text{sp}}$ given by (\ref%
{WouldBeSP}) in the ignorance of the canted phase. We examine how this point
is modified by equating (\ref{NU2gEnergS}) and (\ref{NU2gEnergP}). Namely,
from $\mathcal{E}_{\text{g}}^{\text{spin}}=\mathcal{E}_{\text{var}}^{\text{%
ppin}}$ we find%
\begin{equation}
\frac{\Delta _{\text{SAS}}^{\text{var}}}{E_{\text{C}}^{0}}=\frac{1}{2}\left( 
\frac{\Delta _{\text{SAS}}^{\text{sp}}}{E_{\text{C}}^{0}}-0.1+\sqrt{\left( 
\frac{\Delta _{\text{SAS}}^{\text{sp}}}{E_{\text{C}}^{0}}+0.1\right) ^{2}-%
\frac{d^{2}}{4\ell _{B}^{2}}}\right) .  \label{WouldBeSPx}
\end{equation}%
It is observed that this is much smaller than the first order perturbation
result [FIG.\ref{FigFanBL2SCP}]. Since the spin-canted boundary (\ref%
{NU2PhaseA}) is not modified we expect that the canted phase is shrunk
considerably. These features are precisely what are found in the exact
diagonalization of a few electron system\cite{Schliemann00L} [FIG.\ref%
{FigPD-ZSAS}].

\section{Experimental Status}

\begin{figure}[h]
\includegraphics[width=0.4\textwidth]{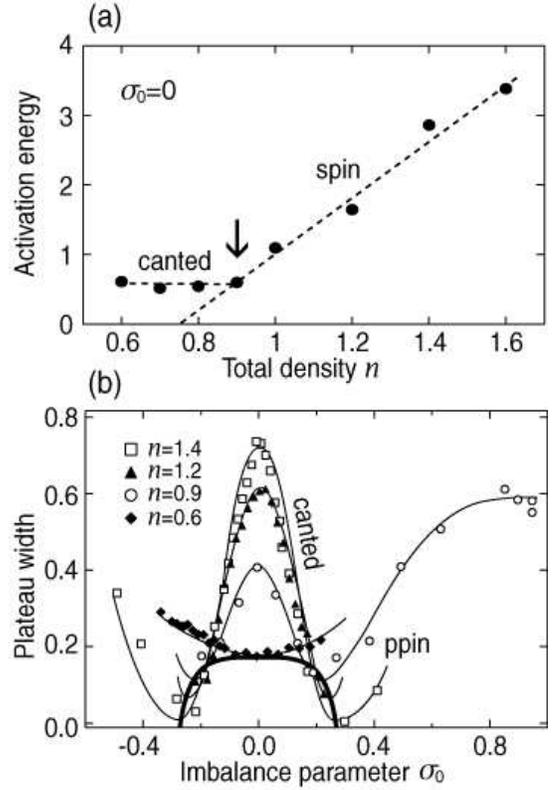}
\caption{The activation-energy data for $\protect\rho _{0}$ and the
plateau-width data for $\protect\sigma _{0}$ are taken from Sawada et al.%
\protect\cite{Sawada98L}. In the upper figure (a), the point $n=0.9$ is
identified with the spin-canted phase transition point. In the lower figure
(b), the minimum point in each curve indexed by $n$ is identified with the
ppin-canted phase transition point. The heavy curve traces these minimum
points.}
\label{FigSawadaPRL}
\end{figure}

Based on our theoretical results we wish to interpret the experimental data
due to Sawada et el.\cite{Sawada98L,Sawada99B} yielding an unambiguous
evidence for phase transitions. They have presented the activation-energy
data for $\rho _{0}$ [FIG.\ref{FigSawadaPRL}(a)] and the plateau-width data
for $\sigma _{0}$ [FIG.\ref{FigSawadaPRL}(b)]. In this section we set $\rho
_{0}=n\times 10^{11}/$cm$^{2}$, and use $n$ to represent the total density.

\begin{figure}[h]
\includegraphics[width=0.4\textwidth]{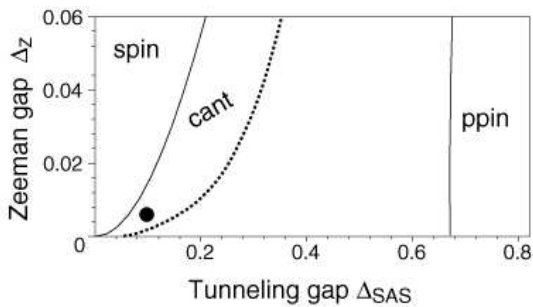}
\caption{The phase diagram is given in the $\Delta _{\text{SAS}}$-$\Delta _{%
\text{Z}}$ plane for $\Delta _{\text{bias}}=0$ and $d=1.23\ell _{B}$. We
have taken the Coulomb energy unit $E_{\text{C}}^{0}$ for the tunneling gap
(horizontal axis) and the Zeeman gap (vertical axis). The large point is the
experimental data indicating a phase transition point taken from Sawada et
al.\protect\cite{Sawada98L}. The two solid curves represent the phase
boundaries in the first order of perturbation. The dotted curve stands for
the ppin-canted phase boundary extrapolated from the exact diagonalization
result obtained at $d=\ell _{B}$ for the 12 electron system\protect\cite%
{Schliemann00L}. It is hard to decide on which phase boundary this point
exists from this phase diagram.}
\label{FigPD-ZSAS123}
\end{figure}

We see a phase transition point $(\sigma _{0},n)=(0,0.9)$ from the
activation-energy data [FIG.\ref{FigSawadaPRL}(a)], for which $d/\ell
_{B}=1.23$ in the sample with $d=23$nm. We give the phase diagram in $\Delta
_{\text{SAS}}$-$\Delta _{\text{Z}}$ plane, upon which we plot the data point
[FIG.\ref{FigPD-ZSAS123}]. On one hand, the point is almost on the
spin-canted boundary. Furthermore, though the spin-canted boundary is exact,
it is obtained in the ideal two-dimensional system and it will be modified
in actual samples with finite quantum wells. Then it is reasonable that the
point is slightly off the theoretical estimation. On the other hand, as we
have argued, the ppin-canted phase boundary is considerably modified so that
the canted phase occupies a tiny domain in the phase diagram [FIG.\ref%
{FigFanBL2SCP} and FIG.\ref{FigPD-ZSAS}]. We cannot compare the exact
diagonalization result\cite{Schliemann00L} with the data directly since it
is available only at $V_{\text{bias}}=0$ and $d/\ell _{B}=1$. Since the
phase diagram for $d/\ell _{B}=1.23$ is similar to the one for $d/\ell
_{B}=1 $, it would be allowed to extrapolate the exact diagonalization data
from $d/\ell _{B}=1$. We have plotted it on the same figure as indicated by
the dotted curve [FIG.\ref{FigPD-ZSAS123}]. Then the point is also near to
the ppin-canted phase boundary. It is hard to decide on which phase boundary
this point exists from this phase diagram.

We answer this problem based on the phase diagram in the $\sigma _{0}$-$\rho
_{0}$ plane [FIG.\ref{FigNU2nPz67}]. We focus on the horizontal line at $%
n=0.9$. This corresponds to the curve indexed by $n=0.9$ in the
plateau-width data [FIG.\ref{FigSawadaPRL}(b)]. The curve takes a maximum at 
$\sigma _{0}=0$ and a minimum at $\sigma _{0}\simeq 0.2$ as the density
imbalance is made. It is reasonable to identify the region $\sigma _{0}>0.2$
with the ppin phase. Then the region $0<\sigma _{0}<0.2$ must be the canted
phase. Note that the spin phase is realized only at $\sigma _{0}=0$. We
conclude that the point $(\sigma _{0},n)=(0.2,0.9)$ represents the
ppin-canted phase transition point and that the point $(\sigma
_{0},n)=(0,0.9)$ found in the activation-energy data represents the
spin-canted phase transition [FIG.\ref{FigNU2nPz67}]. This interpretation is
consistent with the fact that no indication of phase transion is found for $%
0.9<n<1.6$ in the same data [FIG.\ref{FigSawadaPRL}(a)].

We continue to examine the plateau-width data for the sample with various
density $n$ [FIG.\ref{FigSawadaPRL}(b)]. The data with $n=1.4$ and $n=1.2$
show similar behaviors as the one with $n=0.9$. Namely, they possess the
minimum points to be identified with the ppin-canted phase transitions
around the imbalance parameter $\sigma _{0}\simeq 0.3$. It is observed that
the ppin-canted transition point $\sigma _{0}$\ decreases as $n$ decreases.
It is not clear from the data when the ppin-canted phase transition
disappears as $n$ decreases. However, it seems that the entire region
belongs to the ppin phase at $n=0.6$. Let us tentatively regard the critical
point exists near the point $n=0.6$. In this way we have speculated the real
ppin-canted boundary in the phase diagram in the $\sigma _{0}$-$\rho _{0}$
plane [FIG.\ref{FigNU2nPz67}] and in the $\rho _{0}$-$\Delta _{\text{SAS}}$
plane [FIG.\ref{FigNU2DsasN}].

\section{Discussions}

\label{SecDiscus}

There are three phases, i.e., the spin phase, the canted phase and the ppin
phase at $\nu =2$. The ground state is the fully spin polarized state (\ref%
{NU2GrounSpin}) in the spin phase, while it is the fully pseudospin
polarized state (\ref{NU2GrounPpin}) in the ppin phase. In the canted phase,
however, it is not simply made of a certain combination of these two states,
as postulated in a phenomenological bosonic spin model\cite{Demler99L}, but
made of more states in a very complicated way. We have constructed the
ground states explicitly in the first order of perturbation.

The phase diagram has been studied numerically within the Hartree-Fock
approximation in the standard literature\cite%
{Zheng97L,Sarma97L,Brey99L,MacDonald99B}. In this paper we have presented
analytic formulas for it based on perturbation theory. Our results in the
first order approximation reproduce precisely the Hartree-Fock variational
result\cite{MacDonald99B} at the zero bias voltage, and agree qualitatively
with numerical results for nonzero bias voltages\cite{Brey99L,MacDonald99B}.
We have also argued how the phase diagram is modified by higher order
quantum corrections. We have shown that the spin-canted phased boundary is
not modified but the ppin-canted phase boundary is considerably modified. It
is necessary to develop a reliable theory including higher order quantum
corrections to determine the accurate ppin-canted boundary in various phase
diagrams.

Our new contribution is the phase diagrams in the $\sigma _{0}$-$\rho _{0}$
plane as well as in the $\Delta _{\text{bias}}$-$\rho _{0}$ plane. We have
analyzed the relation between the imbalance parameter $\sigma _{0}$ and the
bias voltage $V_{\text{bias}}$. At $\nu =1$ the density imbalance occurs as
soon as the bias voltage becomes nonzero. However, this is not the case at $%
\nu =2$ as in FIG.\ref{FigNU2Pz}, where $\Delta _{\text{bias}}=eV_{\text{bias%
}}$. As far as $V_{\text{bias}}<V_{\text{bias}}^{\text{sc}}$ the density
imbalance is not induced and the system is still in the spin phase. A charge
imbalance occurs only above a certain critical value $V_{\text{bias}}^{\text{%
sc}}$\ of the bias voltage, where the system is driven into the canted
phase. As the bias voltage increases above a second critical value $V_{\text{%
bias}}^{\text{pc}}$, the system turns into the ppin phase. Taking these
facts into account we have constructed the phase diagram in the $\sigma _{0}$%
-$\rho _{0}$ plane as well as in the $\Delta _{\text{bias}}$-$\rho _{0}$
plane. These phase diagrams will be useful to identify the three phases
experimentally by using a single sample. We have also presented the phase
diagram in the $\rho _{0}$-$\Delta _{\text{SAS}}$ plane. It will be useful
to identify the three phases experimentally by using several samples with
different $\Delta _{\text{SAS}}$ but all other parameters unchanged.

We have interpreted experimental data due to Sawada et al.\cite%
{Sawada98L,Sawada99B}. We admit that our analysis is far from satisfactory.
This is mainly because of the lack of qualified data. Recall that the
primary concern of these experiments were to reveal the existence of the
interlayer coherent phase at $\nu =2$, which is identified with the ppin
phase in the present terminology. In passing detailed experiments with
several samples are urged to be performed in order to establish the
ground-state structure in the $\nu =2$ bilayer QH system.

\section{Acknowledgments}

We would like to thank A. Fukuda, Y. Hirayama, S. Kozumi, N. Kumada, K.
Muraki, A. Sawada and D. Terasawa for fruitful discussions on the subject.
We would also like to thank A. MacDonald and J. Schliemann for providing us
with their numerical data of the exact diagonalization result in Ref.\cite%
{Schliemann00L}. ZFE and GT are grateful to the hospitality of Theoretical
Physics Laboratory, RIKEN, where a part of this work was done. ZFE is
supported in part by Grants-in-Aid for Scientific Research from Ministry of
Education, Science, Sports and Culture (Nos. 13135202,14540237); GT
acknowledges a research fellowship from Japan Society for Promotion of
Science (Nos. L04514).

\appendix

\section{Physical Variables}

\label{AppenPV}

The basis of the algebra SU(4) is spanned by 15 Hermitian traceless
matrices. For our purposes it is convenient to choose them as $\tau _{a}^{%
\text{spin}}$, $\tau _{a}^{\text{ppin}}$ and $\tau _{a}^{\text{spin}}\tau
_{b}^{\text{ppin}}$ given by 
\begin{align}
\tau _{a}^{\text{spin}}=& \left( 
\begin{array}{cc}
\tau _{a} & 0 \\ 
0 & \tau _{a}%
\end{array}%
\right) ,\quad \tau _{x}^{\text{ppin}}=\left( 
\begin{array}{cc}
0 & \mathbb{I} \\ 
\mathbb{I} & 0%
\end{array}%
\right) ,  \notag \\
\tau _{y}^{\text{ppin}}=& \left( 
\begin{array}{cc}
0 & -i\mathbb{I} \\ 
+i\mathbb{I} & 0%
\end{array}%
\right) ,\quad \tau _{z}^{\text{ppin}}=\left( 
\begin{array}{cc}
\mathbb{I} & 0 \\ 
0 & -\mathbb{I}%
\end{array}%
\right) ,
\end{align}%
where $\mathbb{I}$ denotes the $2\times 2$ identity matrix, and $\tau _{a}$
are the Pauli matrices.

A basis in the space of $4\times 4$ antisymmetric matrices, comprised of 6
matrices, can be chosen as $\tau _{a}^{\text{spin}}\tau _{x}^{\text{spin}%
}\tau _{y}^{\text{ppin}}$ and $\tau _{a}^{\text{ppin}}\tau _{x}^{\text{spin}%
}\tau _{y}^{\text{ppin}}$. Hence, introducing four tree-dimensional vectors $%
\mathbf{A}$, $\mathbf{B}$, $\mathbf{C}$ and $\mathbf{D}$, we may
parameterize the matrix $g$ as 
\begin{align}
2g=& (A_{x}+iB_{x})\tau _{z}^{\text{spin}}\tau _{y}^{\text{ppin}%
}+(C_{x}+iD_{x})\tau _{y}^{\text{spin}}\tau _{z}^{\text{ppin}}  \notag \\
& +(A_{y}+iB_{y})(-i\tau _{y}^{\text{ppin}})+(C_{y}+iD_{y})(-i\tau _{y}^{%
\text{spin}})  \notag \\
& +(A_{z}+iB_{z})(-\tau _{x}^{\text{spin}}\tau _{y}^{\text{ppin}%
})+(C_{z}+iD_{z})(-\tau _{y}^{\text{spin}}\tau _{x}^{\text{ppin}})  \notag \\
=& -i\left[ (\mathbf{A}+i\mathbf{B})\mathbf{\tau }^{\text{spin}}+(\mathbf{C}%
+i\mathbf{D})\mathbf{\tau }^{\text{ppin}}\right] \tau _{x}^{\text{spin}}\tau
_{y}^{\text{ppin}}.  \label{NU2ggB}
\end{align}%
We come to 
\begin{align}
gg^{\dag }=& \frac{1}{4}(\mathbf{A}^{2}+\mathbf{B}^{2}+\mathbf{C}^{2}+%
\mathbf{D}^{2})  \notag \\
& +\frac{1}{2}\tau _{a}^{\text{spin}}\varepsilon _{abc}A_{b}B_{c}+\frac{1}{2}%
\tau _{a}^{\text{ppin}}\varepsilon _{abc}C_{b}D_{c}  \notag \\
& +\frac{1}{2}\tau _{a}^{\text{spin}}\tau _{b}^{\text{ppin}%
}(A_{a}C_{b}+B_{a}D_{b}).  \label{NU2ggA}
\end{align}%
Comparing this with (\ref{NU2gg}) we get{} 
\begin{subequations}
\label{NU2ConstSPR}
\begin{align}
\mathit{S}_{a}=& \varepsilon _{abc}A_{b}B_{c},  \label{NU2ConstS} \\
\mathit{P}_{a}=& \varepsilon _{abc}C_{b}D_{c},  \label{NU2ConstP} \\
\mathit{R}_{ab}=& A_{a}C_{b}+B_{a}D_{b},  \label{NU2ConstR}
\end{align}%
{}from which we derive the constraints{} 
\end{subequations}
\begin{subequations}
\begin{align}
& \mathit{S}_{a}\mathit{R}_{ab}=0,  \label{NU2ConstA} \\
& \mathit{R}_{ab}\mathit{P}_{b}=0,  \label{NU2ConstB} \\
& \mathit{S}_{a}\mathit{P}_{b}-\varepsilon _{acd}\varepsilon _{bhe}\mathit{R}%
_{ch}\mathit{R}_{de}=0  \label{NU2ConstC}
\end{align}%
{}on $\mathit{S}_{a}$, $\mathit{P}_{a}$ and $\mathit{R}_{ab}$. The
constraint (\ref{NU2ConstC}) is verified based on the identity 
\end{subequations}
\begin{align}
\varepsilon _{acd}\varepsilon _{bhe}=& \delta _{ab}\left( \delta _{ch}\delta
_{de}-\delta _{ce}\delta _{dh}\right) -\delta _{ah}\left( \delta _{cb}\delta
_{de}-\delta _{ce}\delta _{bd}\right)  \notag \\
& +\delta _{ae}\left( \delta _{cb}\delta _{dh}-\delta _{ch}\delta
_{bd}\right) ,
\end{align}%
while (\ref{NU2ConstS}) and (\ref{NU2ConstP}) trivially follow from (\ref%
{NU2ConstA}) and (\ref{NU2ConstB}).

We construct $\mathit{R}_{ab}$ satisfying (\ref{NU2ConstA})--(\ref{NU2ConstC}%
). To satisfy the constraint (\ref{NU2ConstA}) we use the two normalized
vectors%
\begin{equation}
\frac{\mathbf{\mathit{S}}^{2}\mathit{P}_{a}-(\mathbf{\mathit{S}}\mathbf{%
\mathit{P}})\mathit{S}_{a}}{\mathit{S}\mathit{Q}},\qquad \frac{\mathit{Q}_{a}%
}{\mathit{Q}},
\end{equation}%
which are orthogonal to $\mathit{S}_{a}$, and orthogonal one to another. To
satisfy the constraint (\ref{NU2ConstB}) we use the two normalized vectors%
\begin{equation}
\frac{\mathbf{\mathit{P}}^{2}\mathit{S}_{b}-(\mathbf{\mathit{S}}\mathbf{%
\mathit{P}})\mathit{P}_{b}}{\mathit{P}\mathit{Q}},\qquad \frac{\mathit{Q}_{b}%
}{\mathit{Q}},
\end{equation}%
which are orthogonal to $\mathit{P}_{b}$, and orthogonal one to another.
Hence we are able to expand $\mathit{R}_{ab}$ in terms of 4 tensors made out
of these vectors with 4 coefficients $\mathit{R}_{PS}$, $\mathit{R}_{PQ}$, $%
\mathit{R}_{QS}$ and $\mathit{R}_{QQ}$ as in (\ref{NU2StepF}) in text. We
now substitute (\ref{NU2StepF}) into the constraint (\ref{NU2ConstC}) to
find that%
\begin{equation}
\mathit{R}_{QS}\mathit{R}_{PQ}-\mathit{R}_{PS}\mathit{R}_{QQ}=\mathit{S}%
\mathit{P}.  \label{NU2CondiG}
\end{equation}%
Consequently three variables are independent among $\mathit{R}_{PS}$, $%
\mathit{R}_{PQ}$, $\mathit{R}_{QS}$ and $\mathit{R}_{QQ}$. They are
parametrized as in (\ref{NU2StepG}).

\section{Unphysical Variables}

\label{AppenUV}

We have identified 9 independent physical variables in the 12 parameters of
the antisymmetric matrix $g$. In this appendix we identify the 3 unphysical
variables. In so doing we derive the kinematical condition (\ref{NU2CondiM}%
), i.e., $\mathbf{\mathit{S}}^{2}+\mathbf{\mathit{P}}^{2}+\mathbf{\mathit{R}}%
^{2}\leq 1$.

The matrix $g$ is expanded in terms of 4 three-dimensional vectors $\mathbf{A%
}$, $\mathbf{B}$, $\mathbf{C}$ and $\mathbf{D}$ as in (\ref{NU2ggB}). We
investigate how they are given in terms of the physical variables. For this
purpose we reverse the constraints (\ref{NU2ConstS}) and (\ref{NU2ConstP}).
The most general expressions read%
\begin{align}
A_{a}=& \frac{\mathit{S}^{2}\mathit{P}_{a}-(\mathbf{\mathit{S}}\mathbf{%
\mathit{P}})\mathit{S}_{a}}{\mathit{S}\mathit{Q}}\theta _{A}+\frac{\mathit{Q}%
_{a}}{\mathit{Q}}\zeta _{A},  \notag \\
B_{a}=& \frac{\mathit{S}^{2}\mathit{P}_{a}-(\mathbf{\mathit{S}}\mathbf{%
\mathit{P}})\mathit{S}_{a}}{\mathit{S}\mathit{Q}}\theta _{B}+\frac{\mathit{Q}%
_{a}}{\mathit{Q}}\zeta _{B},  \notag \\
C_{a}=& \frac{\mathit{P}^{2}\mathit{S}_{a}-(\mathbf{\mathit{S}}\mathbf{%
\mathit{P}})\mathit{P}_{a}}{\mathit{P}\mathit{Q}}\theta _{C}+\frac{\mathit{Q}%
_{a}}{\mathit{Q}}\zeta _{C},  \notag \\
D_{a}=& \frac{\mathit{P}^{2}\mathit{S}_{a}-(\mathbf{\mathit{S}}\mathbf{%
\mathit{P}})\mathit{P}_{a}}{\mathit{P}\mathit{Q}}\theta _{D}+\frac{\mathit{Q}%
_{a}}{\mathit{Q}}\zeta _{D},  \label{NU2ConstD}
\end{align}%
where the parameters $\theta _{A,B}$ and $\zeta _{A,B}$ are restricted by 
\begin{align}
\theta _{A}\zeta _{B}-\zeta _{A}\theta _{B}& =\mathit{S},  \label{NU2CondiD}
\\
\theta _{D}\zeta _{C}-\zeta _{D}\theta _{C}& =\mathit{P}.  \label{NU2CondiF}
\end{align}%
We substitute (\ref{NU2ConstD}) into (\ref{NU2ConstR}), and compare the
resulting equation with (\ref{NU2StepF}). In this way we obtain new
restrictions, 
\begin{align}
\left[ 
\begin{array}{cc}
\theta _{A} & \theta _{B} \\ 
\zeta _{A} & \zeta _{B}%
\end{array}%
\right] \left[ 
\begin{array}{c}
\theta _{C} \\ 
\theta _{D}%
\end{array}%
\right] =& \left[ 
\begin{array}{c}
\mathit{R}_{PS} \\ 
\mathit{R}_{QS}%
\end{array}%
\right] ,  \notag \\
\left[ 
\begin{array}{cc}
\theta _{A} & \theta _{B} \\ 
\zeta _{A} & \zeta _{B}%
\end{array}%
\right] \left[ 
\begin{array}{c}
\zeta _{C} \\ 
\zeta _{D}%
\end{array}%
\right] =& \left[ 
\begin{array}{c}
\mathit{R}_{PQ} \\ 
\mathit{R}_{QQ}%
\end{array}%
\right] .
\end{align}%
We solve these as 
\begin{align}
\theta _{C}=& \frac{\mathit{R}_{PS}\zeta _{B}-\mathit{R}_{QS}\theta _{B}}{%
\mathit{S}},\quad \theta _{D}=\frac{\mathit{R}_{QS}\theta _{A}-\mathit{R}%
_{PS}\zeta _{A}}{\mathit{S}},  \notag \\
\zeta _{C}=& \frac{\mathit{R}_{PQ}\zeta _{B}-\mathit{R}_{QQ}\theta _{B}}{%
\mathit{S}},\quad \zeta _{D}=\frac{\mathit{R}_{QQ}\theta _{A}-\mathit{R}%
_{PQ}\zeta _{A}}{\mathit{S}}.  \label{NU2ConstE}
\end{align}%
It is trivial to check that (\ref{NU2CondiF}) is automatically satisfied due
to (\ref{NU2CondiG}) and (\ref{NU2CondiD}). We count the number of
independent variables. There are 9 physical variables; $\mathit{S}_{a}$, $%
\mathit{P}_{a}$, $\mathit{R}_{PS}$, $\mathit{R}_{PQ}$, $\mathit{R}_{QS}$ and 
$\mathit{R}_{QQ}$ with one constraint (\ref{NU2CondiG}). There are 3 extra
variables; $\theta _{A}$, $\theta _{B}$, $\zeta _{A}$ and $\zeta _{B}$ with
one constraint (\ref{NU2CondiD}). We now show that they are the unphysical
variables.

A well-known unphysical variable is the overall phase of the ground state.
First we identify it. We consider two vectors $\left( \theta _{A},\theta
_{B}\right) $ and $\left( \zeta _{A},\zeta _{B}\right) $. We rotate the two
vectors $\left( \theta _{A},\theta _{B}\right) $ and $\left( \zeta
_{A},\zeta _{B}\right) $ by a single angle $\phi $, which is the overall
angle common to them. It turns out that the two vectors $\left( \theta
_{C},\theta _{D}\right) $ and $\left( \zeta _{C},\zeta _{D}\right) $ rotate
by the same angle due to (\ref{NU2ConstE}). Then, $\left( A_{a},B_{a}\right) 
$ and $\left( C_{a},D_{a}\right) $ rotate in the same way due to (\ref%
{NU2ConstB}). Finally, (\ref{NU2ggB}) implies that the matrix $g$ acquires
the angle $\phi $, which is the overall phase of the ground state.

Another well-known unphysical variable is associated with the normalization
of the ground state. The normalization condition (\ref{NU2GrounNorma}) is
equivalent to 
\begin{equation}
\mathbf{A}^{2}+\mathbf{B}^{2}+\mathbf{C}^{2}+\mathbf{D}^{2}=2.
\label{NU2CondiO}
\end{equation}%
Substituting (\ref{NU2ConstD}) into this and using (\ref{NU2ConstE}), we
obtain 
\begin{align}
\mathit{S}^{2}=& \frac{1}{2}\left( \theta _{A}^{2}+\theta _{B}^{2}\right)
\left( \mathit{S}^{2}+\mathit{R}_{QS}^{2}+\mathit{R}_{QQ}^{2}\right)  \notag
\\
+& \frac{1}{2}\left( \zeta _{A}^{2}+\zeta _{B}^{2}\right) \left( \mathit{S}%
^{2}+\mathit{R}_{PS}^{2}+\mathit{R}_{PQ}^{2}\right)  \notag \\
-& \left( \theta _{A}\zeta _{A}+\theta _{B}\zeta _{B}\right) (\mathit{R}_{QS}%
\mathit{R}_{PS}+\mathit{R}_{QQ}\mathit{R}_{PQ}).  \label{NU2CondiH}
\end{align}%
We make a change of variables. Let $\gamma $ be the angle between the
vectors $\left( \theta _{A},\theta _{B}\right) $ and $\left( \zeta
_{A},\zeta _{B}\right) $. Omitting the overall angle we may write 
\begin{align}
\left( \theta _{A},\theta _{B}\right) =& \left( \theta \mathrm{\cos }\frac{%
\gamma }{2},-\theta \mathrm{\sin }\frac{\gamma }{2}\right) ,  \notag \\
\left( \zeta _{A},\zeta _{B}\right) =& \left( \zeta \mathrm{\cos }\frac{%
\gamma }{2},\zeta \mathrm{\sin }\frac{\gamma }{2}\right) .  \label{NU2CondiP}
\end{align}%
Next, we introduce new variables $x$ and $y$ by 
\begin{align}
x& =\theta ^{2}\left[ 1+\frac{\mathit{R}_{QS}^{2}+\mathit{R}_{QQ}^{2}}{%
\mathit{S}^{2}}\right] +\zeta ^{2}\left[ 1+\frac{\mathit{R}_{PS}^{2}+\mathit{%
R}_{PQ}^{2}}{\mathit{S}^{2}}\right] ,  \notag \\
y& =\theta ^{2}\left[ 1+\frac{\mathit{R}_{QS}^{2}+\mathit{R}_{QQ}^{2}}{%
\mathit{S}^{2}}\right] -\zeta ^{2}\left[ 1+\frac{\mathit{R}_{PS}^{2}+\mathit{%
R}_{PQ}^{2}}{\mathit{S}^{2}}\right] ,  \label{NU2CondiI}
\end{align}%
which we use instead of $\theta ^{2}$ and $\zeta ^{2}$. Now, there are three
variables $x$, $y$ and $\gamma $ with two constraints (\ref{NU2CondiD}) and (%
\ref{NU2CondiH}).

We denote 
\begin{equation}
a=\frac{\mathit{S}^{2}\left( \mathit{S}^{2}+\mathit{P}^{2}+\mathit{R}%
^{2}\right) }{\left( \mathit{S}^{2}+\mathit{R}_{QS}^{2}+\mathit{R}%
_{QQ}^{2}\right) \left( \mathit{S}^{2}+\mathit{R}_{PS}^{2}+\mathit{R}%
_{PQ}^{2}\right) },  \label{NU2CondiJ}
\end{equation}%
or%
\begin{equation}
\frac{a}{1-a}=\frac{\mathit{S}^{2}\left( \mathit{S}^{2}+\mathit{P}^{2}+%
\mathit{R}^{2}\right) }{(\mathit{R}_{QS}\mathit{R}_{PS}+\mathit{R}_{QQ}%
\mathit{R}_{PQ})^{2}}.
\end{equation}%
Here 
\begin{equation}
\mathit{R}^{2}\equiv \mathit{R}_{ab}^{2}=\mathit{R}_{PS}^{2}+\mathit{R}%
_{PQ}^{2}+\mathit{R}_{QS}^{2}+\mathit{R}_{QQ}^{2},
\end{equation}%
where the last equality follows from (\ref{NU2StepF}).

Using the above notations two constraints (\ref{NU2CondiD}) and (\ref%
{NU2CondiH}) are rearranged into 
\begin{align}
\mathrm{\cos }\gamma =& \frac{x-2}{\sqrt{(1-a)(x^{2}-y^{2})}},  \notag \\
\mathrm{\sin }\gamma =& \frac{2\sqrt{\mathit{S}^{2}+\mathit{P}^{2}+\mathit{R}%
^{2}}}{\sqrt{a(x^{2}-y^{2})}}.  \label{NU2CondiN}
\end{align}%
These are well defined since $x^{2}>y^{2}$ and $0<a<1$, as follows from
definitions (\ref{NU2CondiI}) and (\ref{NU2CondiJ}). It follows that 
\begin{equation}
\frac{(x-2)^{2}}{(1-a)(x^{2}-y^{2})}+\frac{4(\mathit{S}^{2}+\mathit{P}^{2}+%
\mathit{R}^{2})}{a(x^{2}-y^{2})}=1
\end{equation}%
from $\mathrm{\cos }^{2}\gamma +\mathrm{\sin }^{2}\gamma =1$, or%
\begin{equation}
\left( ax-2\right) ^{2}+a\left( 1-a\right) y^{2}=4\left( 1-a\right) \left( 1-%
\mathit{S}^{2}-\mathit{P}^{2}-\mathit{R}^{2}\right) .  \label{NU2CondiK}
\end{equation}%
We ask the question whether (\ref{NU2CondiK}) admits solutions for $x$ and $%
y $ which leads to positive values of $\theta ^{2}$ and $\zeta ^{2}$ via (%
\ref{NU2CondiI}). The solvability condition turns out to be 
\begin{equation}
\mathit{S}_{a}^{2}+\mathit{P}_{a}^{2}+\mathit{R}_{ab}^{2}\leqslant 1.
\label{NU2CondiL}
\end{equation}%
This is the condition (\ref{NU2CondiM}) for the magnitude of the isospin.

If the condition is satisfied, we may solve $x$ and $y$ from (\ref{NU2CondiK}%
), and obtain $\theta $, $\zeta $ and $\gamma $ from (\ref{NU2CondiI}) and (%
\ref{NU2CondiN}). Solutions are not unique. Eq.(\ref{NU2CondiK}) defines an
ellipse. Moving along the ellipse the parameters $\theta $, $\zeta $ and $%
\gamma $ take different values, and therefore lead to different vectors $%
\mathbf{A}$, $\mathbf{B}$, $\mathbf{C}$ and $\mathbf{D}$. However, all these
vectors lead to one and the same set of physical fields $\mathit{S}_{a}$, $%
\mathit{P}_{a}$ and $\mathit{R}_{ab}$. This is the unphysical variable
inherent to the $\nu =2$ bilayer QH system, which is commented below (\ref%
{NU2StepG}). This unphysical mode decouples on the ground state since the
ellipse\ is shrunk to a point, as we see in the following appendix.

\section{Ground-State Condition}

\label{AppenGSC}

The eigenvalue equation (\ref{NU2CondiGrounA}) in the SU(4)-invariant system
leads to the condition (\ref{NU2CondiGrounB}), or%
\begin{equation}
\epsilon _{\alpha \beta \mu \nu }g_{\alpha \beta }g_{\mu \nu }=0.
\label{AppNCa}
\end{equation}%
On the other hand, the variational ground state in the full
SU(4)-noninvariant system is found to satisfy (\ref{NU2CondiA}), or%
\begin{equation}
\mathbf{\mathit{S}}^{2}+\mathbf{\mathit{P}}^{2}+\mathbf{\mathit{R}}^{2}=1.
\label{AppNCb}
\end{equation}%
We verify the equivalence of these two conditions.

For this purpose, we substitute (\ref{NU2ggB}) into (\ref{AppNCa}), and use (%
\ref{NU2CondiO}), to obtain%
\begin{equation}
\mathbf{A}^{2}+\mathbf{D}^{2}=\mathbf{B}^{2}+\mathbf{C}^{2}=1,\quad \mathbf{%
AB}=\mathbf{CD}.  \label{AppNCc}
\end{equation}%
On the other hand, examining this step of transformation we see that (\ref%
{AppNCa}) follows from (\ref{AppNCc}). Namely, (\ref{AppNCa}) is equivalent
to (\ref{AppNCc}) based on the formula (\ref{NU2ggB}).

The derivation of (\ref{AppNCb}) from (\ref{AppNCc}) is easy. Using (\ref%
{NU2ConstSPR}) we express $\mathbf{\mathit{S}}^{2}+\mathbf{\mathit{P}}^{2}+%
\mathbf{\mathit{R}}^{2}$ in terms of $\mathbf{A}$, $\mathbf{B}$, $\mathbf{C}$
and $\mathbf{D}$. We then use (\ref{AppNCc}) to derive (\ref{AppNCb}).

The derivation of (\ref{AppNCc}) from (\ref{AppNCb}) is more complicated.
When the condition (\ref{AppNCb}) holds, the ellipse defined by (\ref%
{NU2CondiK}) is shrunk to a point set given by $ax=2$ and $y=0$. In this
case, $\theta $ and $\zeta $ are determined by solving (\ref{NU2CondiI}), 
\begin{equation}
\theta =\sqrt{\mathit{S}^{2}+\mathit{R}_{PS}^{2}+_{PQ}^{2}},\quad \zeta =%
\sqrt{\mathit{S}^{2}+\mathit{R}_{QS}^{2}+\mathit{R}_{QQ}^{2}},
\label{AppNCd}
\end{equation}%
and (\ref{NU2CondiN}) becomes 
\begin{equation}
\cos \gamma =\sqrt{1-a},\qquad \sin \gamma =\sqrt{a},  \label{AppNCe}
\end{equation}%
where $a$ is given by (\ref{NU2CondiJ}). First we note that%
\begin{equation}
\mathbf{A}^{2}+\mathbf{D}^{2}=\theta _{A}^{2}+\zeta _{A}^{2}+\theta
_{D}^{2}+\zeta _{D}^{2},
\end{equation}%
to which we substitute (\ref{NU2ConstE}) and (\ref{NU2CondiP}). We then use (%
\ref{NU2CondiH}), (\ref{AppNCd}) and (\ref{AppNCe}). In this way we prove $%
\mathbf{A}^{2}+\mathbf{D}^{2}=1$ after some straightforward calculation.
Similarly we can verify all equations in (\ref{AppNCc}).

As we have seen, the unphysical variables $\theta $, $\zeta $ and $\gamma $
are fixed in terms of the physical variables on the ground state, where the
condition (\ref{AppNCb}) is satisfied. Thus, every ground-state
configuration has a unique matrix $g_{\mu \nu }$. We substitute (\ref%
{NU2StepG}), (\ref{NU2StepJ}) and (\ref{NU2StepA}) into (\ref{AppNCd}) and (%
\ref{AppNCe}). Choosing $\omega =\pi /4$, for simplicity, we find 
\begin{equation}
\theta =\zeta =\frac{\sqrt{\Delta _{0}^{2}+\Delta _{Z}^{2}\left( 1-\beta
^{2}\right) }}{\Delta _{0}}\sqrt{1-\alpha ^{2}},  \label{NU2EllipC}
\end{equation}%
and 
\begin{align}
\cos \frac{\gamma }{2}=& \frac{\Delta _{0}}{\sqrt{\Delta _{0}^{2}+\Delta
_{Z}^{2}\left( 1-\beta ^{2}\right) }},  \notag \\
\sin \frac{\gamma }{2}=& \frac{\Delta _{Z}\sqrt{1-\beta ^{2}}}{\sqrt{\Delta
_{0}^{2}+\Delta _{Z}^{2}\left( 1-\beta ^{2}\right) }}.  \label{NU2EllipD}
\end{align}%
Using (\ref{NU2EllipC}) and (\ref{NU2EllipD}) in (\ref{NU2CondiP}) we obtain 
\begin{align}
& \theta _{A}=\zeta _{A}=\sqrt{1-\alpha ^{2}},  \notag \\
& \theta _{B}=-\zeta _{B}=-\frac{\Delta _{Z}}{\Delta _{0}}\sqrt{1-\alpha ^{2}%
}\sqrt{1-\beta ^{2}}.
\end{align}%
Now, it is straightforward to express $\mathbf{A}$, $\mathbf{B}$, $\mathbf{C}
$ and $\mathbf{D}$ in terms of $\alpha $ and $\beta $, and hence to
reconstruct the matrix $g_{\mu \nu }$ as in (\ref{NU2StepK}).


\begin{thebibliography}{99}
\bibitem{BookEzawa} Z.F. Ezawa, \textit{Quantum Hall Effects: Field
Theoretical Approach and Related Topics} (World Scientific, 2000).

\bibitem{BookDasSarma} S. Das Sarma and A. Pinczuk (eds), \textit{%
Perspectives in Quantum Hall Effects} (Wiley, 1997).

\bibitem{NoteA} In the standard literature the spin-ferromagnet and
pseudospin-singlet phase is called the ferromagnet phase, while the
spin-singlet and pseudospin-ferromagnet phase is called the singlet phase.
Here, we call them the spin phase and the ppin phase to respect their equal
partnership.

\bibitem{Zheng97L} L. Zheng, R.J. Radtke and S. Das Sarma, Phys. Rev. Lett.
78 (1997) 2453.

\bibitem{Sarma97L} S. Das Sarma, S. Sachdev and L. Zheng, Phys. Rev. Lett.
79 (1997) 917; Phys. Rev. B58 (1998) 4672.

\bibitem{Pellegrini97L} V. Pellegrini, A. Pinczuk, B.S. Dennis, A.S. Plaut,
L.N. Pfeiffer and K.W. West, Phys. Rev. Lett. 78 (1997) 310; Science 281
(1998) 799.

\bibitem{Sawada98L} A. Sawada, Z.F.Ezawa, H. Ohno, Y. Horikoshi, Y. Ohno, S.
Kishimoto, F. Matsukura, M. Yasumoto and A. Urayama Phys. Rev. Lett. 80
(1998) 4534.

\bibitem{Brey99L} L. Brey, E. Demler and S. Das Sarma, Phys. Rev. Lett. 83
(1999) 168.

\bibitem{Schliemann00L} J. Schliemann and A.H. MacDonald, Phys. Rev. Lett.
84 (2000) 4437.

\bibitem{EzawaX04B} Z.F. Ezawa and G. Tsitsishvili, Phys. Rev. B70 (2004)
125304.

\bibitem{Hasebe02B} K. Hasebe and Z.F. Ezawa, Phys. Rev. B66 (2002) 155318.

\bibitem{EzawaX03B} Z.F. Ezawa, G. Tsitsishvili and K. Hasebe, Phys. Rev.
B67 (2003) 125314.

\bibitem{MacDonald99B} A.H. MacDonald, R. Rajaraman and T. Jungwirth, Phys.
Rev. B60 (1999) 8817.

\bibitem{Sawada99B} A. Sawada, Z.F. Ezawa, H. Ohno, Y. Horikoshi, A.
Urayama, Y. Ohno, S. Kishimoto, F. Matsukura and N. Kumada, Phys. Rev. B59
(1999) 14888.

\bibitem{Demler99L} E. Demler and S. Das Sarma, Phys. Rev. Lett. 82 (1999)
3895.
\end{thebibliography}
\end{document}